\documentclass[a4paper,11pt]{article}
\pdfoutput=1 

\usepackage{jcappub}
\usepackage[T1]{fontenc}
\usepackage{centernot}

\usepackage{array}
\usepackage{tensor}
\usepackage{amsthm}
\usepackage{amsmath}
\usepackage{amssymb}
\usepackage{chngcntr}
\usepackage{commath}
\usepackage{caption}
\usepackage{graphicx}
\usepackage{graphics}
\usepackage{physics}
\usepackage{mathtools}
\usepackage{multirow}
\usepackage{float}
\usepackage{subfiles}
\usepackage[dvipsnames]{xcolor}
\usepackage{epstopdf}

\usepackage[normalem]{ulem}

\newtheorem{theorem}{Theorem}[section]
\newtheorem{proposition}[theorem]{Proposition}

\newcommand{\sums}{\sum\limits}
\newcommand{\I}{{(i)}}
\newcommand{\J}{{(j)}}
\newcommand{\munu}{{\mu\nu}}
\newcommand{\bigcups}{\bigcup\limits}
\newcommand{\m}{\mathcal{M}}
\newcommand{\s}{\mathcal{S}}
\newcommand{\0}{{(0)}}
\newcommand{\1}{{(1)}}
\newcommand{\2}{{(2)}}
\newcommand{\3}{{(3)}}
\newcommand{\4}{{(4)}}

\newcommand{\p}{{(p)}}
\newcommand{\q}{{(q)}}
\newcommand{\opn}{\mathcal{O}_{PN}}
\newcommand{\oa}{\mathcal{O}_{\alpha}}
\newcommand{\eff}{{\textrm{eff}}}

\title{Parameterised post-Newtonian formalism for the effective field theory of dark energy via screened reconstructed Horndeski theories}
\author[a, b]{Cyril Renevey,}
\author[a]{Joe Kennedy,}
\author[a]{Lucas Lombriser}

\affiliation[a]{D\'{e}partement de Physique Th\'{e}orique, Universit\'{e} de Gen\`{e}ve, \\ 24 quai Ernest Ansermet, 1211 Gen\`{e}ve 4, Switzerland}
\affiliation[b]{Institute for Theoretical Physics, ETH Z\"urich, \\ Wolfgang-Pauli-Strasse 27, 8093 Z\"urich, Switzerland}

\emailAdd{renevey@lpsc.in2p3.fr, Joseph.Kennedy@unige.ch, Lucas.Lombriser@unige.ch}

\abstract{
We bring together two popular formalisms which generically parameterise deviations from General Relativity on astrophysical and cosmological scales, namely the parameterised post-Newtonian (PPN) formalism and the effective field theory (EFT) of dark energy and modified gravity.
These separate formalisms are successfully applied to independently perform tests of gravity in their respective regimes of applicability on vastly different length scales.
Nonlinear screening mechanisms indeed make it imperative to probe General Relativity across a wide range of scales.
For a comprehensive interpretation of the complementary measurements it is important to connect them to effectively constrain the vast gravitational model space.
We establish such a connection within the framework of Horndeski scalar-tensor theories restricted to a luminal propagation speed of gravitational waves. This is possible via the reconstruction of the family of linearly degenerate covariant Horndeski actions from the set of EFT functions and the subsequent derivation of the PPN parameters from the reconstructed theory.
We outline the required conditions which ensure a reconstructed Horndeski model possesses a screening mechanism that enables significant modifications on cosmological scales while respecting stringent astrophysical bounds.
Employing a scaling method, we then perform the general post-Newtonian expansion of the reconstructed models to derive their PPN parameters $\gamma$ and $\beta$ in their screened regimes.
}

\begin{document}

\maketitle
\flushbottom

\section{Introduction} \label{sec:intro}

General Relativity (GR) has successfully passed all tests from a wealth of high-precision astrophysical measurements to date~\cite{Stairs:2003eg,Will:2014kxa,Manchester:2015mda,Archibald:2018oxs,Will:2018bme,Renevey:2019jrm,Baker:2019gxo}. 
New Solar System and pulsar experiments such as the FAST and SKA radio telescopes
will further tighten these bounds~\cite{Shao:2018qpt,Berge:2019phz,Pan:2020zsw}. 
In parallel to the astrophysical probes, complementary efforts are being made to perform precise tests of GR on cosmological scales~\cite{Copeland:2006wr,Clifton:2011jh,Koyama:2015vza,Joyce:2016vqv,Ishak:2018his}. 
During the past two decades there has been a significant improvement in cosmological tests of gravity with the increasing quantity and quality of data. 
In upcoming years we will benefit from further high-precision experiments that will enable us to place more stringent constraints on GR. 
While the astrophysical constraints are generally far tighter than the cosmological bounds, it is important to emphasise the vastly different length scales involved. 
Considering the relevant orders of magnitude, the difference in scale between astrophysical and cosmological tests is comparable to that between the diameter of the atomic nucleus and the realm of everyday human experience.    
It is therefore imperative to perform independent tests of GR in the cosmological regime. 
This necessity is further emphasised by the emergence of nonlinear screening mechanisms~\cite{Vainshtein:1972sx,Damour:1994zq,Khoury:2003rn,Babichev:2009ee,Hinterbichler:2011ca} (see Refs.~\cite{Joyce:2014kja,Joyce:2016vqv} for reviews) in modified theories of gravity that suppress deviations from GR for astrophysical probes.

Traditionally, a particularly important driver for the development of modified gravity theories on cosmological scales has been the evidence for the late-time accelerated expansion of the Universe~\cite{Riess:1998cb,Perlmutter:1998np}. 
While the cosmological constant offers the simplest explanation, a number of theoretical obstacles~\cite{Weinberg:1988cp,Padilla:2015aaa,Martin:2012bt,Burgess:2013ara} (cf.~\cite{Kaloper:2013zca,Wang:2017oiy,Lombriser:2019jia}) have led theorists to study a variety of alternative scenarios to explain the underlying mechanism behind cosmic acceleration~\cite{Joyce:2014kja, Joyce:2016vqv}. 
The recent confirmation of a luminal speed of gravitational waves~\cite{TheLIGOScientific:2017qsa,Lombriser:2015sxa} challenged modified gravity as the direct cause of the acceleration~\cite{Lombriser:2016yzn}.  
Nevertheless, the poorly understood dominating dark sector provides sufficient motivation for performing thorough tests of gravity.
For instance, a dark energy component may still couple non-minimally to matter and, while this interaction would not give rise to cosmic acceleration, the resulting modification of gravity can still leave an observable impact on cosmological scales while being suppressed on astrophysical scales.

The sheer size of the gravitational model space strongly motivates the development of systematic approaches to comprehensively explore their cosmological and astrophysical implications. 
Consequently, a great effort has gone into developing generalised and efficient parameterisation frameworks for testing gravity (see Ref.~\cite{Lombriser:2018guo} for a review) that enable one to effectively discriminate between gravitational models with observational data.

The parameterised post-Newtonian (PPN) formalism~\cite{Will:1971zza} fulfils this purpose for astrophysical phenomena subject to a low-energy static approximation. 
In this limit, one can perform a post-Newtonian (PN) expansion in which the metric is expanded in the ratio of its velocities to the speed of light, neglecting the background evolution and assuming an asymptotic Minkowskian limit. 
The PPN formalism parameterises a generic PN expansion with ten parameters in terms of the PN order of the modification~\cite{Will:1971zza}, enabling stringent model-independent tests of gravity. 
Owing to screening mechanisms, these constraints cannot straightforwardly be applied to cosmological scales. 
Conversely, although cosmological modifications of gravity may be endowed with effective screening mechanisms, deviations from GR may not vanish completely from astrophysical systems. 
Importantly, the inherently nonlinear nature of screening mechanisms complicates the application of PPN to models exhibiting screening~\cite{Avilez-Lopez:2015dja,Hohmann:2015kra,Zhang:2016njn,McManus:2017itv,Bolis:2018kcq}. 
This can be attributed to the linearisation performed in the PN expansion which renders screening ineffective. 
Crucially, the PPN formalism is also not suitable for evolving backgrounds and thus for cosmology. 
Note that extensions of the PPN formalism to cosmological scales have been pursued by patching together PN expansions in small regions of spacetime to obtain a PPN cosmology~\cite{Sanghai:2016tbi,Clifton:2018cef}, or by performing a post-Friedmannian expansion of the cosmological metric in powers of the inverse light speed~\cite{Milillo:2015cva,Thomas:2020duj}.

Inspired by the successes of PPN, the development of an equivalent formalism for tests of gravity on cosmological scales has been the subject of intensive research (see Ref.~\cite{Lombriser:2016yzn} for a review). 
Effective field theory (EFT)~\cite{Creminelli:2008wc,Park:2010cw,Bloomfield:2011np,Gubitosi:2012hu,Bloomfield:2012ff,Gleyzes:2013ooa,Gleyzes:2014rba,Bellini:2014fua,Lombriser:2015cla,Lagos:2016wyv,Lombriser:2018olq,Frusciante:2019xia} has proved to be particularly practical for the generalised description of the cosmological background and linear perturbations in gravitational theories with additional degrees of freedom (also see Refs.~\cite{Cusin:2017mzw,Frusciante:2017nfr} for extensions to nonlinear perturbations). 
An EFT action is constructed by summing operators which are consistent with the symmetries imposed on the system up to a specified order.
Extensive efforts are devoted to constrain the EFT parameter space with current and upcoming cosmological surveys such as Euclid~\cite{Laureijs:2011gra} and LSST~\cite{Ivezic:2008fe} (e.g., Refs.~\cite{SpurioMancini:2019rxy,Noller:2018wyv,Frusciante:2018jzw}).
For a comprehensive interpretation of observational constraints on gravity it is necessary to connect cosmological and astrophysical formalisms so that bounds inferred from data in either regime can be combined. 
It is indeed important to study the observational implications of the model space across a wide range of length scales. 
For example, this point was emphasised in Ref.~\cite{Kennedy:2019nie} where it was demonstrated that scalar-tensor theories cannot exhaustively be distinguished from $\Lambda$CDM up to arbitrary finite $n$-th order in the cosmological perturbations. 
Astrophysical constraints are therefore needed to complement cosmological constraints and vice-versa.
%

%
In Ref.~\cite{Lombriser:2018guo} it was proposed that a link between the two formalisms can be established within the framework of general scalar-tensor theories. 
Scalar-tensor theories are among the most popular extensions to GR~\cite{Copeland:2006wr,Clifton:2011jh,Koyama:2015vza,Joyce:2016vqv,Ishak:2018his}, introducing an additional scalar degree of freedom in the Einstein-Hilbert action that modifies the gravitational dynamics. 
This addition may lead to instabilities associated with higher-order equations of motion~\cite{Woodard:2006nt} and so it is necessary to restrict the structure of the theory in such a way that they can be avoided. 
This is accomplished with the Horndeski action~\cite{Horndeski:1974wa, Deffayet:2011gz, Kobayashi:2011nu} which represents the most general set of covariant scalar-tensor theories with second-order equations of motion. 
Note however that it is possible to construct stable scalar-tensor theories with higher-order equations of motion~\cite{Gleyzes:2014dya, Langlois:2015cwa}. %
Due to the essentially infinite freedom in Horndeski theory, the EFT of dark energy is tremendously useful to describe its phenomenology without reference to a specific model. 
All of the effects of Horndeski theory on the cosmological background and linear perturbations can then be encoded in just five EFT parameters. 
Their functional form is determined either by a specific Horndeski theory or chosen with a phenomenological motivation without reference to a particular underlying model. 
Conversely, starting from EFT, one may use a reconstruction method~\cite{Kennedy:2017sof,Kennedy:2018gtx} that maps from a set of EFT functions to the set of Horndeski theories which are degenerate at the level of the background and linear perturbations. This reconstruction can also be extended to $n$-th order perturbations in EFT~\cite{Kennedy:2019nie}. 
As pointed out in Ref.~\cite{Lombriser:2018guo}, this reconstruction can be used to recover the family of covariant Horndeski models for a given set of EFT functions, from which then the PPN parameters can be inferred, hence establishing a link between the two formalisms.

While the linear cosmological behaviour of Horndeski theories is exhaustively described by the EFT of dark energy, their PPN description has not yet been fully developed~\cite{Avilez-Lopez:2015dja,Hohmann:2015kra,Zhang:2016njn,McManus:2017itv,Bolis:2018kcq}. 
This is predominantly due to the complexity arising from the nonlinear screening mechanisms that are active on astrophysical scales. 
These are an essential component of scalar-tensor theories and as aforementioned allow significant modifications of GR at cosmological scales while still enabling them to satisfy stringent astrophysical constraints. 
Developing the PPN formalism for Horndeski theories is therefore an important aim of current research. 
Recently, progress has been made for the subclasses of Galileon interactions~\cite{Avilez-Lopez:2015dja,McManus:2017itv,Bolis:2018kcq} and chameleon gravity~\cite{McManus:2017itv}. 
The scaling method of Ref.~\cite{McManus:2016kxu} has proven particularly useful to describe the screening properties of general Horndeski theories, enabling one to perform the PN expansion in screened regimes.

\begin{figure}
    \centering
    \includegraphics[width=0.6\textwidth]{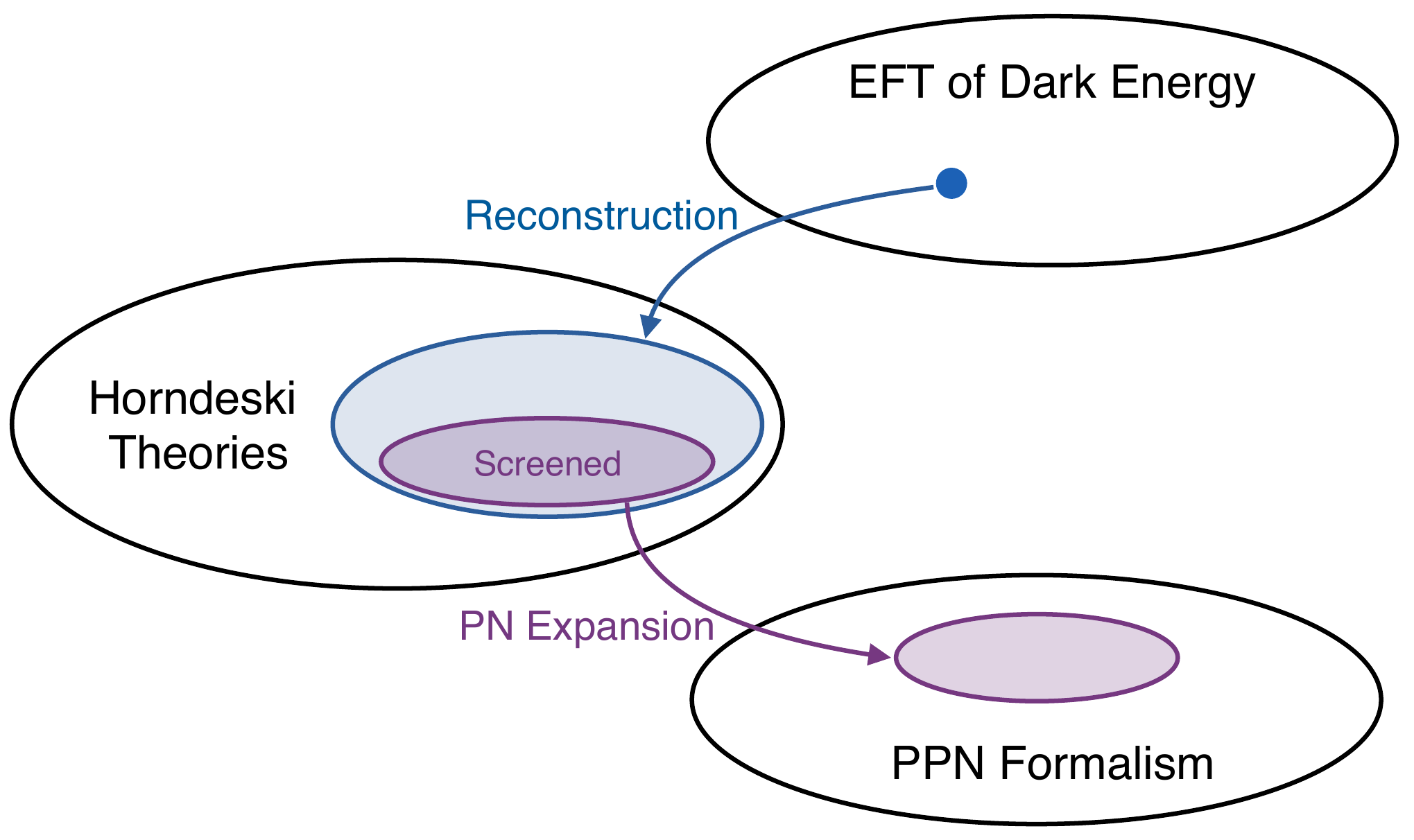}
    \caption{
    %
    %
    %
    After mapping to the set of Horndeski theories with equivalent EFT expansions using the reconstruction method of Ref.~\cite{Kennedy:2017sof} we then identify the subset of these theories that possess a screening mechanism.  
    By performing a PN expansion of these screened reconstructed theories we can then derive the resultant PPN parameters in terms of the original EFT parameters.
    A reconstructed covariant theory from EFT can thus help bridge the gap between cosmological and astrophysical tests of Horndeski theory. 
    %
    }
    \label{fig:roadmap}
\end{figure}

In this paper we will pursue two aims. 
We will first develop the PPN formalism for general Horndeski theories employing the scaling method under the restriction of a luminal propagation speed of tensor modes as motivated by the recent gravitational wave measurements~\cite{TheLIGOScientific:2017qsa,Lombriser:2015sxa}. 
The formalism should cover all known classes of screening mechanisms. 
Secondly, we apply the formalism to the reconstructed Horndeski models from the EFT of dark energy and modified gravity on cosmological scales, thus establishing a connection between the cosmological EFT and astrophysical PPN formalisms in the framework of Horndeski scalar-tensor gravity. An outline of the paper is presented in Fig.~\ref{fig:roadmap}. 

The paper is organised as follows. 
In Sec.~\ref{sec:background} we briefly review
the theoretical ingredients we will utilise to connect the PPN and EFT formalisms. 
We discuss Horndeski theories, the EFT of dark energy and modified gravity on cosmological scales, and the reconstruction method of mapping from the EFT framework to covariant Horndeski theories.
We also summarise the main aspects of the PPN formalism for astrophysical tests of gravity that are relevant for scalar-tensor theories and the scaling method that enables a PN expansion in screened regimes. 
In Sec.~\ref{sec:einstein_gravity_limit} we employ the scaling method to construct a technique that tests
whether a reconstructed Horndeski theory with $c_T=1$ possesses a screening mechanism or conversely can be used to exploit the nonlinear freedom of reconstructed models to incorporate a screening mechanism. 
The connection between the PPN and EFT formalisms within the framework of Horndeski gravity is then developed in Sec.~\ref{sec:PPN},
where we perform the PN expansion of the reconstructed theories and compare it to the standard PPN expansion to derive the PPN parameters.
We conclude with a summary of our results in Sec.~\ref{sec:conclusions}.
Finally, in the appendix we provide some useful relations employed in the derivations presented in Secs.~\ref{sec:background}--\ref{sec:PPN}, we develop the scaling method at the level of the action, 
and we provide an example of how our formalism can be used to infer parameter constraints with pulsar systems.

\section{Parameterising Horndeski models for astrophysics and cosmology} \label{sec:background}

Before developing the connection between the PPN formalism and the EFT of dark energy in the framework of Horndeski gravity in Sec.~\ref{sec:PPN}, we shall first briefly introduce the necessary theoretical background for establishing this link. 
In Sec.~\ref{subsubsec:horndeski} we present a brief review of Horndeski gravity and discuss the different classes of screening mechanisms that can operate in it.
In Sec.~\ref{sec:cosmologicalscales} we review some aspects of the effective field theory of dark energy that parameterises Horndeski modifications for linear cosmological perturbations.
EFT does not capture the inherently nonlinear screening effects. 
For this purpose we discuss a reconstruction method which maps from a set of EFT parameters to a family of linearly degenerate Horndeski models, providing the necessary nonlinear completion to study screening.
For slowly evolving weak-field gravitational phenomena the cosmological EFT formalism becomes inadequate, and in Sec.~\ref{sec:astrophysicalscales} we briefly review the parameterised post-Newtonian formalism appropriate for this regime.
However, PPN is inadequate for cosmological applications due to its static assumption.
As with EFT, the linearisation performed in the PPN expansion complicates the straightforward description of screening effects. W
We therefore discuss a scaling method which enables a consistent expansion in screened regimes.
We will use these tools in Sec.~\ref{sec:einstein_gravity_limit} to develop an expansion in the screening limits of Horndeski theory before connecting the PPN and EFT formalisms in Sec.~\ref{sec:PPN}.

\subsection{Horndeski gravity}\label{subsubsec:horndeski}\label{subsec:Horndeski}

We restrict our study to scalar-tensor gravity.
The most general four-dimensional, local, Lorentz-covariant scalar-tensor theory with second-order equations of motion is given by the Horndeski action~\cite{Horndeski:1974wa}
\begin{align}
S_{\emph{H}}\big[\phi,g\big] =  & \frac{M_p^2}{2} \int \dd^4x \sqrt{-g} \bigg\lbrace G_2(\phi,X) - G_3(\phi,X) \Box \phi \nonumber \\
&+G_4(\phi,X) R + G_{4X}(\phi,X) [ (\Box \phi)^2 - (\nabla_\mu \nabla_\nu\phi)^2 ] \nonumber \\
& +G_5(\phi,X) G_{\mu \nu}\nabla^\mu \nabla^\nu \phi - \frac{G_{5X}(\phi,X)}{6}[(\Box \phi)^3 - 3\Box \phi(\nabla_\mu \nabla_\nu \phi)^2 + 2(\nabla_\mu \nabla_\nu \phi)^3] \bigg\rbrace \nonumber\\
&+ S_m[g] \,,\label{eq:horndeski_action}
\end{align}
where $R$ is the Ricci scalar of the metric $g_\munu$, $X\equiv-\frac{1}{2}\nabla_\mu\phi\nabla^\mu\phi$ is the kinetic term of the scalar field, $G_i(\phi,X)$ with $i=2,...,5$ are arbitrary functions, $M_p^2=(8\pi G)^{-1}$ denotes the reduced Planck mass squared, and we have set the speed of light in vacuum to $c=1$.
The subscripts $\phi$ and $X$ in $G_{i\phi}$ or $G_{iX}$ represent partial derivatives of the function with respect to $\phi$ or $X$ respectively. 
Motivated by the recent upper bound set on the propagation speed of tensor modes $(c_T-1)<10^{-15}$~\cite{TheLIGOScientific:2017qsa,Lombriser:2015sxa} we will assume $c_T=1$ in this work. 
This bound effectively sets $G_{4X}=G_5=0$ in the Horndeski action~\cite{McManus:2016kxu}, although novel mechanisms that can avoid this constraint are presented in Refs.~\cite{Oikonomou:2020sij, Battye:2018ssx, deRham:2018red, Copeland:2018yuh}. 
We assume $G_{4X}=G_5=0$ for the rest of this work unless otherwise stated.
The equations of motion which follow from the action in Eq.~\eqref{eq:horndeski_action} are given by~\cite{Kobayashi:2011nu, McManus:2016kxu}
\begin{align}
    &G_4(\phi) R_\munu=-\sums_{i=2}^4R_\munu^\I+\frac{1}{M_p^2}\left(T_\munu-\frac{1}{2}g_\munu T\right)\,,\label{eq:MFE_horndeski}\\
    &\sums_{i=2}^4\left(\nabla^\mu J_\mu^\I-P_\phi^\I\right)G_4(\phi)+\sums_{i=2}^4R^\I=-\frac{T}{M_p^2} \,,\label{eq:SFE_horndeski}
\end{align}
where $T_\munu$ is the energy-momentum tensor, $R_\munu^\I$, $J_\mu^\I$, and $P_\phi^\I$ are defined in Appendix \ref{app:field_equations}, and $R^\I$ is the trace of $R_\munu^\I$. Solar System and astrophysical tests of gravity have already placed tight constraints on the presence of a non-minimally coupled extra scalar degree of freedom in high-curvature regions~\cite{Will:2018bme,Stairs:2003eg}.
Effects of a scalar field modifying gravity at cosmological scales should therefore be highly suppressed in these regimes.
This requirement has led to the
intensive study of screening mechanisms 
that
suppress modifications of GR in the Solar System and for most astrophysical objects.
Many scalar-tensor theories possess a screening mechanism which can be classified into three main categories~\cite{Joyce:2016vqv}.
(i) Large field value screening is effective in regions where the Newtonian potential $\Phi_N$ is larger than a certain threshold $\Lambda$. 
Examples of this type of screening are the chameleon~\cite{Khoury:2003rn}, symmetron~\cite{Hinterbichler:2011ca} and dilaton~\cite{Damour:1994zq} mechanisms. 
(ii) Screening via first derivatives of the scalar field takes place in strong acceleration regimes where $\nabla \Phi_N>\Lambda^2$, which generally corresponds to $\partial \phi/\Lambda^2\gg 1$. 
A typical example is k-mouflage~\cite{Babichev:2009ee}.
(iii) Second derivative screening suppresses modifications in high curvature regions where $\grad^2 \Phi_N >\Lambda^3$, 
which usually translates into the condition $\Box\phi/\Lambda^3\gg 1$ for the scalar field. 
The Vainshtein mechanism is a good example of this class of screening~\cite{Vainshtein:1972sx}.

\subsection{Cosmological scales} \label{sec:cosmologicalscales}

Rather than parameterising the function $G_i(\phi, X)$ in Eq.~\eqref{eq:SFE_horndeski} as arbitrary functions of $\phi$ and $X$, it is often more practical to parameterise the effective gravitational modifications they induce in a given regime of interest.
On linear cosmological scales, the effective field theory of dark energy and modified gravity~\cite{Gubitosi:2012hu,Bloomfield:2012ff,Creminelli:2008wc,Park:2010cw,Bloomfield:2011np,Gleyzes:2013ooa,Gleyzes:2014rba,Bellini:2014fua,Lombriser:2015cla,Lagos:2016wyv,Lombriser:2018olq} has proven to be especially useful.
For the unfamiliar reader, in Sec.~\ref{subsec:EFT} we will briefly review the aspects of the EFT formalism that are of particular relevance to this work.
In Sec.~\ref{subsec:reconstruction} we will then review how the family of covariant Horndeski theories that correspond to a given set of EFT parameters can be recovered with the reconstruction method of Refs.~\cite{Kennedy:2017sof,Kennedy:2018gtx}.

\subsubsection{Effective field theory of dark energy and modified gravity}\label{subsec:EFT}

In developing the EFT of dark energy and modified gravity one begins by foliating spacetime with space-like hypersurfaces as in the Arnowitt-Deser-Misner (ADM) formulation of GR~\cite{Arnowitt:1962hi}, described by the line element
\begin{align}
    \dd s^2=-N^2\dd t^2+h_{ij}(\dd x^i+N^i\dd t)(\dd x^j+N^j\dd t) \,,
\end{align}
where $N$ is the lapse, $N^i$ the shift, and $h_{ij}$ is the induced metric on the three-dimensional hypersurfaces. 
In this formalism diffeomorphism invariance allows us to choose a foliation of spacetime such that the scalar field is completely determined by the time coordinate $\phi=\phi(t)$. 
This choice is called the unitary gauge and as a consequence the scalar field perturbations do not appear explicitly in the action, rather they are absorbed into the time-time component of the metric $g^{00}$. 
For simplicity we choose the relation between the scalar field and the time coordinate to be
\begin{align}
   \phi(t)=t M_*^2 \,,\label{eq:phi_t} 
\end{align}
where $M_*$ is the bare Planck mass. 
The normal vector to each space-like hypersurface is given by
\begin{align}
    n_\mu:=-\frac{\nabla_\mu\phi}{\sqrt{-(\partial\phi)^2}}=-\frac{\delta^0_\mu}{\sqrt{-g^{00}}} \, ,
\end{align}
the intrinsic curvature of the hypersurface is given by the three-dimensional Ricci tensor $R^{(3)}_{\mu\nu}$, and the extrinsic curvature tensor is denoted $K_{\mu\nu}$.   

Full diffeomorphism invariance is lost in the unitary gauge yet it can be restored using the Stückelberg trick
by shifting the time coordinate by a small perturbation $\pi(x)$ as $t\rightarrow t+\pi(x)$. 
The scalar field perturbation $\pi(x)$ is then explicitly re-introduced into the action as the Nambu-Goldstone boson of the spontaneously broken time translation symmetry.
Furthermore, with the definition in Eq.~\eqref{eq:phi_t} we see that the kinetic term can be written as
\begin{align}
    X=-\frac{1}{2}M_*^4g^{00}=-\frac{1}{2}M_*^4(-1+\delta g^{00})\, ,\label{eq:X_delta_g00}
\end{align}
where $\delta g^{00}$ is the perturbation of the 00-component of the metric.
 
The EFT action describing the dynamics of both the cosmological background and linear perturbations of Horndeski theory is then given by \cite{Gubitosi:2012hu, Bloomfield:2012ff, Gleyzes:2013ooa}  
\begin{align}
S =& S^{(0,1)}+S^{(2)}\label{eq:EFT_action}\\
S^{(0,1)} = &\frac{M_{*}^{2}}{2}\int d^{4}x \sqrt{-g} \left[ \Omega(t) R -2\Lambda(t)-\Gamma(t)\delta g^{00} \right] \, ,
 \\
S^{(2)} = & \int d^{4}x \sqrt{-g} \bigg[ \frac{1}{2}M^{4}_2(t)(\delta g^{00})^2-\frac{1}{2}\bar{M}^{3}_{1}(t) \delta K \delta g^{00} \nonumber\\
 & -\bar{M}^{2}_{2}(t) \left( \delta K^2-\delta  K^{\mu\nu} \delta K_{\mu\nu} - \frac{1}{2} \delta R^{(3)}\delta g^{00} \right) \bigg] \, ,
\end{align}
where the time dependent functions are called the EFT parameters, $\delta R^\3$ is the perturbation of the three-dimensional Ricci scalar of the constant time hypersurfaces, and we have adopted the notation of Refs.~\cite{Lombriser:2014ira,Kennedy:2017sof}.
The functions $\Omega(t)$, $\Lambda(t)$ and $\Gamma(t)$ are related by the background Friedmann equations 
\begin{align}
    \Gamma&+\Lambda=3\left(\Omega H^2+\dot{\Omega}H\right)-\frac{\rho_m}{M_*^2}\label{eq:friedman1} \,,\\
    \Lambda&=2\Omega \dot{H}+3\Omega H^2+2\dot{\Omega} H+\ddot{\Omega} \,,\label{eq:friedman2}
\end{align}
where $H$ is the Hubble parameter,
$\rho_m$ is the matter density and dots denote time derivatives. 
The set of seven EFT functions
\begin{align}
    \left\{\Omega(t),\Lambda(t),\Gamma(t),M_{2}^{4}(t),\bar{M}_{1}^{3}(t),\bar{M}_{2}^{2}(t),H(t)\right\} \label{eq:set_omega}
\end{align} 
can then be reduced to five free functions with Eqs.~\eqref{eq:friedman1} and \eqref{eq:friedman2}.
It is then possible to express these EFT functions in different bases.
For example Ref.~\cite{Bellini:2014fua} expressed them as the following physically motivated set,
\begin{align}
    \left\{\alpha_M(t),\alpha_B(t),\alpha_K(t),\alpha_T(t),H(t)\right\} \,,\label{eq:set_alpha}
\end{align}
which are related to the basis in Eq.~\eqref{eq:set_omega} as
\begin{align}
    \alpha_{M} & = \frac{M_{*}^{2}\Omega^{\prime}+2(\bar{M}_{2}^{2})^{\prime}}{M_{*}^{2}\Omega+2\bar{M}_{2}^{2}} \,,& 
\alpha_{B} & = \frac{M_{*}^{2}H\Omega^{\prime}+\bar{M}_{1}^{3}}{2H\left( M_{*}^{2}\Omega+2\bar{M}_{2}^{2}     \right)} \,, \nonumber\\
\alpha_{K} & = \frac{M_{*}^{2}\Gamma+4M_{2}^{4}}{H^{2}\left(M_{*}^{2}\Omega+2\bar{M}_{2}^{2} \right)} \,, & 
\alpha_{T} & = -\frac{2\bar{M}_{2}^{2}}{M_{*}^{2}\Omega+2\bar{M}_{2}^{2}} \,, \label{eq:alpha_EFT}
\end{align}
where primes indicate derivatives with respect to $\ln a$ with $a$ being the scale factor. 
The parameter $\alpha_M$ denotes the evolution rate of the squared effective Planck mass $M^2(t)$, $\alpha_B$ quantifies the coupling between the metric and the scalar field, $\alpha_K$ is a coefficient of the kinetic term of the scalar field, and $\alpha_T$ describes the deviation of the speed of gravitational waves from the speed of light, where $\alpha_T=0$ is assumed throughout this paper (Sec.~\ref{subsubsec:horndeski}). 
Alternatively, Refs.~\cite{Kennedy:2018gtx,Lombriser:2018olq} formulated another basis which inherently avoids gradient and ghost instabilities.

Note that the linear expansion performed with the EFT formalism precludes from capturing the inherently nonlinear screening effects of a model.
In principle, a linear shielding mechanism can operate in Horndeski theories though this ruled out by the $\alpha_T=0$ constraint~\cite{Lombriser:2014ira,Lombriser:2015sxa}.
In order to describe the nonlinear behaviour associated with a set of EFT parameters, one must reconstruct the family of covariant Horndeski Lagrangians that gives rise to the original set of EFT parameters.

\subsubsection{Reconstructed Horndeski models}\label{subsec:reconstruction}

\begin{table}[t]
\centering
\renewcommand{\arraystretch}{2}
\begin{tabular}{|c|c|} 
\hline
$\quad U(\phi) = \Lambda + \frac{\Gamma}{2}-\frac{M_{2}^{4}}{2M_{*}^{2}}- \frac{9H\bar{M}_{1}^{3}}{8M_{*}^{2}} - \frac{(\bar{M}_{1}^{3})^{\prime}}{8}\quad$&$\quad b_{1}(\phi)=\frac{\bar{M}_{1}^{3}}{2M_{*}^{6}}\quad $\\ 
\hline
$a_{2}(\phi)=\frac{M_{2}^{4}}{2M_{*}^{8}}+\frac{(\bar{M}^{3}_{1})^{\prime}}{8M_{*}^{6}}-\frac{3H\bar{M}_{1}^{3}}{8M_{*}^{8}}$ & $F(\phi)=\Omega$     \\ \hline 
\multicolumn{2}{|c|}
{$Z(\phi) = \frac{\Gamma}{M_{*}^{4}} - \frac{2M_{2}^{4}}{M_{*}^{6}} - \frac{3H\bar{M}_{1}^{3}}{2M_{*}^{6}} + \frac{(\bar{M}_{1}^{3})^{\prime}}{2M_{*}^{4}}$}      \\ 
\hline 
\end{tabular}
\renewcommand{\arraystretch}{1}
\captionof{table}{Coefficients of the Horndeski functions $G_i(\phi,X)$ expanded in $X$ in Eqs.~\eqref{eq:rec1}--\eqref{eq:rec3}, reconstructed from the cosmological EFT parameters.} 
\label{table:mapping_EFT_to_horndeski}
\end{table}

Starting from a particular Horndeski action, one can express it in the unitary gauge, perform an expansion in the perturbations, and derive the corresponding EFT parameters~\cite{Bloomfield:2013efa, Bellini:2014fua}. 
Conversely, Refs.~\cite{Kennedy:2017sof, Kennedy:2018gtx} developed the reverse mapping from a given set of EFT parameters to the class of Horndeski theories that are degenerate at the level of the cosmological background and linear perturbations.
%
The reconstructed family of Horndeski theories is given by
\begin{align}
    G_{2}(\phi, X) = & -M_{*}^{2}U(\phi)+M_{*}^{2} Z(\phi)X+4a_{2}(\phi)X^{2}+\Delta G_{2} \,, \label{eq:rec1} \\[0.5em]
G_{3}(\phi,X) = &-2b_{1}(\phi)X+\Delta G_{3} \,, \label{eq:rec2} \\
G_{4}(\phi, X) = & \: \frac{1}{2}M_{*}^{2}F(\phi) \,, \label{eq:rec3}
\end{align}
where the functions $U$, $Z$, $a_2$, $b_1$ and $F$ are defined in terms of the EFT parameters in Table~\ref{table:mapping_EFT_to_horndeski}. 
The terms $\Delta G_{i}$ can be added to the reconstructed theory without affecting the background and linear perturbations and are specified by 
\begin{align}
    \Delta G_i(\phi,X)=\sums_{n \geq 3}\xi_n^\I(\phi)\left(1-\frac{2X}{M_*^4}\right)^n \, ,
    \label{eq:correction_terms}
\end{align}
where $\xi_n^\I(\phi)$ are arbitrary functions.
Eq.~\eqref{eq:correction_terms} characterises the degenerate family of Horndeski theories that yield the same set of EFT parameters.
Note that the form of Eq.~\eqref{eq:correction_terms} differs from Ref.~\cite{Kennedy:2017sof} due to the different definition of $X$.

%
%
%
%

In Sec.~\ref{sec:einstein_gravity_limit} it will prove useful to write the Horndeski functions as a polynomial in $X/M_*^4$ instead of $1-2X/M_*^4$. 
Assuming each $\Delta G_{i}$ to be a finite polynomial we then obtain
\begin{align}
    \Delta G_i(\phi, X)=\sums_{n= 3}^{N^\I} \sums_{m=0}^n {{n}\choose{m}} \left( \frac{-2X}{M_*^4} \right)^m\xi^\I_n(\phi) \,,\label{eq:DGi}
\end{align}
where $N^\I$ is the order of the polynomial. 
Using the definitions in Table~\ref{table:xi_n}
one can express each term in the reconstructed action as a single sum given by
\begin{align}
    G_i(\phi, X)=\sums_{m=0}^{N^\I} \sums_{n=m}^{N^\I} {{n}\choose{m}} \left(\frac{-2X}{M_*^4}\right)^m\xi^\I_n(\phi) \,.\label{eq:DGi2}
\end{align}
This expansion describes the set of reconstructed Horndeski theories with $c_T=1$ and we will henceforth work 
with $G_i$ functions of this form.
Finally, note that it was demonstrated in Ref.~\cite{Kennedy:2019nie} that the reconstruction method can be extended to higher orders in EFT.
It is therefore possible to express each function at $n$-th order in perturbations $\xi^\I_n(\phi)$ in terms of a set of nonlinear EFT parameters.
Constraints on these higher-order EFT contributions would determine the functional form of each $\xi^\I_n(\phi)$ thus removing the reconstructed theory's nonlinear degeneracy order-by-order in the regime where one can perform a perturbative expansion. 
In addition, the expansion can be used to implement a screening mechanism in the reconstructed model~\cite{Kennedy:2019nie} (Sec.~\ref{sec:einstein_gravity_limit}).

\begin{table}[]
\centering
\renewcommand{\arraystretch}{1.6}
\begin{tabular}{|c|c|}
\hline
$\xi_0^{(2)}(\phi)=-M_*^2U(\phi)+\frac{1}{2}M_*^6Z(\phi)+M_*^8a_2(\phi)$ & {$\xi_1^{(2)}(\phi)=-\frac{1}{2}M_*^6Z(\phi)-2M_*^8a_2(\phi)$}                            \\ \hline
$\xi_2^{(2)}(\phi)=M_*^8a_2(\phi)$                                       & {$\xi_0^{(3)}(\phi)=-M_*^4b_1(\phi)$}                                                     \\ \hline
$\xi_1^{(3)}(\phi)=M_*^4b_1(\phi)$                                       & {$\xi_2^{(3)}=0$}                              \\ \hline
\multicolumn{2}{|c|}{$\xi_0^{(4)}(\phi)=\frac{1}{2}M_*^2F(\phi)$} \\ \hline
\end{tabular}
\renewcommand{\arraystretch}{1}
\captionof{table}{Definition of the functions $\xi_n^\I(\phi)$ at the lowest orders in the EFT expansion when writing the Horndeski functions as an expansion in $(1-2X/M_*^4)$ as in Eq.~\eqref{eq:correction_terms}. The analogous result with the EFT functions $\{\alpha_M(t),\alpha_B(t),\alpha_K(t)\}$ can be found in Appendix~\ref{app:reconstruction}.\label{table:xi_n}}
\end{table}

\subsection{Astrophysical scales} \label{sec:astrophysicalscales}

Generic tests of gravity have successfully been conducted with slowly evolving weak-field gravitational phenomena that are well described by the low-energy static limit of GR, a regime particularly applicable to Solar System tests of gravity~\cite{Will:2014kxa}.
In this limit one can perform a post-Newtonian expansion in which the metric is expanded in orders of $(v/c)$, neglecting cosmological evolution and assuming an asymptotic Minkowski limit.
The expansion can be parameterised for generalisations of gravitational interactions, enabling model-independent tests of gravity.
For readers unfamiliar with the parameterised post-Newtonian formalism, we shall present a brief summary of its most relevant aspects to this work in Sec.~\ref{subsubsec:PPN}.
Given the neglect of the background evolution, the PPN formalism is not suitable for cosmology (see, however, Refs.~\cite{Sanghai:2016tbi, Clifton:2018cef, Milillo:2015cva,Thomas:2020duj} for generalisations).
Importantly, because of the linearisation of the field equations in the post-Newtonian expansion, the nonlinear interactions that give rise to screening mechanisms are removed.
Obtaining a direct mapping from the screened models to the PPN formalism and the straightforward comparison to observational parameter bounds is therefore not straightforward. 
Screening effects can also depend on ambient density, giving rise to both low-energy limits where screening operates and where it does not. In Sec.~\ref{subsubsec:scaling_method} we briefly review the scaling method of Refs.~\cite{McManus:2016kxu,McManus:2017itv} that has been developed to deal with these complications and enable a post-Newtonian expansion in screened regimes.

\subsubsection{Parameterised post-Newtonian formalism}
\label{subsubsec:PPN}

The PPN formalism was developed in Ref.~\cite{Will:1971zza} to 
describe deviations from the PN expansion of GR in terms of ten additional parameters. 
The ten PPN parameters characterising the formalism are chosen such that they describe violations of the strong equivalence principle (SEP).
The SEP states that the freely falling motion of self-gravitating bodies is independent of their composition and any local systems satisfy Lorentz and position invariance \cite{Will:2018bme}.
Only two of the PPN parameters in scalar-tensor theories deviate from their GR values, namely $\gamma$, which describes spacetime curvature produced by a unit rest mass, and $\beta$, which accounts for non-linearities in the superposition law of gravity~\cite{Will:2014kxa}. 
These two parameters affect the freely falling motion of self-gravitating bodies so that objects with different densities will behave differently in the same gravitational field.
For a complete description of the PPN formalism see Ref.~\cite{Will:2018bme}.
The order of the PPN expansion is set by the velocity $v\sim\opn(1)$ 
and from the virial relation \cite{Straumann:2013spu} we can relate this to the Schwarzschild radius with $R_s\sim v^2\sim \opn(2)$. 
One can show \cite{Will:2018bme} that 
the PN expansion of
the metric $g_\munu:=\eta_\munu+h_\munu$ is given by
\begin{subequations}
\begin{align}
    g_{00}&=-1+h_{00}^\2+h_{00}^\4+\opn(6) \, , \\
    g_{0i}&=h_{0i}^\1+h_{0i}^\3+\opn(5) \, , \\
    g_{ij}&=+1+h_{ij}^\2+\opn(4) \, ,
\end{align}
\label{eq:PN_metric}%
\end{subequations}
where $h_\munu^\I\sim \opn(i)$ are the correction terms to the Minkowski metric. 
The $0i$-components of the metric do not contribute to $\gamma$ and $\beta$ and so we omit them in this work.
We represent matter as a perfect fluid with matter density $\rho$, pressure $p$ and specific energy density $\Pi$. 
The matter density as well as the specific energy density are of the order of the Newtonian potential $\Phi_N$, namely $\rho\sim\Pi\sim \Phi_N\sim\opn(2)$ and the pressure is comparable to the gravitational energy $p\sim\rho \Phi_N\sim\opn(4)$. 
The energy-momentum tensor $T_\munu$ can then be written up to order $\opn(4)$ as~\cite{Will:2018bme}
\begin{subequations}
\begin{align}
    T_{00}&=\rho\left(1+\Pi+v^2-h_{00}^\2\right)+\opn(6) \,,\\[0.1cm]
    T_{ij}&=\rho v^iv^j+p\delta^{ij}+\opn(6) \,.
\end{align}\label{eq:PN_Tmunu}%
\end{subequations}
In GR, the PN metric can be found by solving the Einstein field equations at order $\opn(4)$ with the approximations \eqref{eq:PN_metric} and \eqref{eq:PN_Tmunu}.
Adding the contributions from the parameters $\gamma$ and $\beta$ to the GR solution yields the PPN metric 
\begin{subequations}
\begin{align}
    g_{00}=&-1+2\Phi_N-2\beta \Phi_N^2+(2\gamma+2)\Phi_1\nonumber\\&+2(-2\beta+1)\Phi_2+2\Phi_3+6\gamma\Phi_4+\opn(6) \,, \\
    g_{ij}=&\ (1+2\gamma \Phi_N)\delta_{ij}+\opn(4) \,,
\end{align}
\label{eq:PPN_metric}%
\end{subequations}
where the potentials $\Phi_N$ and $\Phi_i$, $i=1,...,4$ are defined in Appendix~\ref{app:NP}. 

The goal of this work is to derive the PN expansion of reconstructed Horndeski theories in their screened region and compare it to the PPN metric in Eq.~\eqref{eq:PPN_metric} to obtain a prediction for the PPN parameters.
This will establish a connection between the EFT and PPN formalisms in the framework of Horndeski gravity (Sec.~\ref{sec:PPN}).

\subsubsection{Scaling method for the expansion in screened regimes}\label{subsubsec:scaling_method}

Since screening mechanisms are inherently nonlinear, the linearisation of the field equations performed in the post-Newtonian expansion prevents straightforward identification of the PPN parameters for Horndeski models endowed with screening mechanisms.
There are the models of particular interest for cosmological modifications of gravity.
Ref.~\cite{McManus:2016kxu} proposed a universal technique called the scaling method to both test whether a scalar-tensor theory possesses a screening mechanism, as well as enable an expansion of the model in its screened regime.

In order to apply the scaling method, we start by expanding the scalar field 
in terms of a field perturbation $\psi$ as
\begin{align}
    \phi=\phi_0(1+\alpha^q\psi) \,,
\end{align}
where $\phi_0$ denotes the background value of the scalar field and $\alpha$ is a theoretical coupling constant raised to an arbitrary power $q\in \mathbb{R}$. 
To ensure the perturbation around the background field value $\phi_0$ is linear we require $\alpha^q\psi\ll 1$.
One can now apply this expansion to the scalar field equation, which schematically becomes 
\begin{align}
    \alpha^{s+mq}F_1(\psi,\tilde{X})+\alpha^{t+nq}F_2(\psi,\tilde{X})=\frac{T}{M_p^2} \,,\label{eq:SFE_scaling_method}
\end{align}
with $n,m\in \mathbb{N}$, $s,t\in\mathbb{
R}$ and $\tilde{X}=-\frac{1}{2}\nabla^\mu\psi\nabla_\mu\psi$. 
In general, the left-hand side of Eq.~\eqref{eq:SFE_scaling_method} can have additional terms with different powers of $\alpha$ which we disregard here for the sake of simplicity. 
The main feature of the scaling method is to set an appropriate value for $q$ such that in the limit $\alpha\rightarrow \infty$ or $\alpha\rightarrow 0$, we are left with the field equations in the desired regime, where $\psi\ll \alpha$ or $\psi\gg\alpha$ respectively.
Returning to the example in Eq.~\eqref{eq:SFE_scaling_method} we can assume without loss of generality that $-s/m<-t/n$. 
Since the right-hand side of Eq.~\eqref{eq:SFE_scaling_method} is independent of $\alpha$ it is necessary for at least one term to be independent of $\alpha$ on the left-hand side which survives in the limit. 
Hence $q$ can take two possible values, namely $q\in\{-s/m,-t/n\}$. 
With $q=-s/m$ the limit $\alpha\rightarrow 0$ is divergent, but the limit $\alpha\rightarrow \infty$ yields the field equation
\begin{align}
    F_1(\psi,\tilde{X})=\frac{T}{M_p^2} \,.
\end{align}
Alternatively, with the choice of $q=-t/n$ the limit $\alpha\rightarrow 0$ gives the scalar field equation in the opposite regime
\begin{align}
    F_2(\psi,\tilde{X})=\frac{T}{M_p^2} \,.
\end{align}
If one generalises the concept to a left-hand side of the form $\sum \alpha^{s_i+m_iq}F_i$ with $i=1,...,N$ and $s_i\in\mathbb{R}$ and $m_i\in\mathbb{N}^*$, then the set of possible values for $q$ is $Q:=\{-s_i/m_i\}_i$. 
In the regime where $\alpha\rightarrow\infty$ we need to choose $q=\min Q$, whereas in the limit $\alpha\rightarrow 0$ we take $q=\max Q$.

It is also necessary to determine the form of the metric field equation in the same limit.  
In particular, the field equations should converge with the chosen $q$-value in the required limit.
If this is not the case the theory is not well defined in this regime. 
Furthermore, the Einstein field equations should be recovered in the screened region, shown to be $\alpha\rightarrow\infty$ for derivative screening and $\alpha\rightarrow 0$ for large field value screening~\cite{McManus:2016kxu}. 

The scaling method therefore acts to identify and extract the dominant terms in the equations of motion in a given limit such as a high-curvature regime.
Once can then perform an expansion in $\alpha$, which is either valid in the screened or unscreened regime (Sec.~\ref{sec:einstein_gravity_limit}).

\section{Screening reconstructed Horndeski theories} \label{sec:einstein_gravity_limit}

As we have seen in Sec.~\ref{sec:background}
the class of reconstructed Horndeski theories from a set of EFT functions is degenerate with respect to nonlinear correction terms. 
This degeneracy allows us to choose a reconstructed theory with predetermined cosmological behaviour that exhibits a screening mechanism. 
Because of its screening property, such a theory is more likely to satisfy stringent astrophysical constraints \cite{Will:2014kxa, Renevey:2019jrm}.
We shall now develop a technique to test for, and if needed incorporate, a screening mechanism in a reconstructed theory of the form~\eqref{eq:DGi2}. 
We restrict to Horndeski theories with $c_T=1$ motivated by the LIGO constraint on the propagation speed of tensor modes \cite{TheLIGOScientific:2017qsa}. 
This ensures $G_4$ is solely a function of the scalar field and we
make use of the fact that one can always
redefine the scalar field such that
$G_4(\phi)=\phi$.
We shall focus on the screened limit.
The procedure to find the unscreened limit of reconstructed theories following the method developed in Ref.~\cite{McManus:2016kxu} will be described in Appendix~\ref{app:unscreened}.

We organise this section as follows. 
In Sec.~\ref{subsec:Exp_alpha_Gi} we find the $\alpha$-dependence of the Horndeski functions written as in Eq.~\eqref{eq:DGi2} after the scalar field expansion $\phi=\phi_0(1+\alpha^q\psi)$. 
In Sec.~\ref{subsec:Einstein_limit_luminal} we develop the testing method for an Einstein gravity limit on reconstructed theories with luminal propagation speed of gravitational waves. 
In particular, we will distinguish the cases of large field value and derivative screening. 
We illustrate the testing method in Sec.~\ref{subsec:examples} by adding different screening mechanisms to a reconstructed theory, namely the chameleon, k-mouflage, and Vainshtein mechanisms. 
Finally, we note that while the scaling method will be applied at the level of the equations of motion as in Refs.~\cite{McManus:2016kxu,McManus:2017itv}, in Appendix~\ref{subsec:Einstein_limit_all} we show that it can also be applied directly at the level of the action.

\subsection{Expansion of the Horndeski functions
}\label{subsec:Exp_alpha_Gi}

In order to apply the scaling method to reconstructed Horndeski theories we first need to understand how to extract the $\alpha$-dependence from the Horndeski functions. 
As a first step we wrote the functions $G_i$, $i=2,3$ as an expansion in $(X/M_p^4)^m$.  
With the set $\{\xi_n^\I(\phi)\}_{n\leq 2}$ defined in Table~\ref{table:xi_n} we then wrote the Horndeski functions as in Eq.~\eqref{eq:DGi2},
which shall serve as our starting point.
%
Each $\xi^\I_n(\phi)$ depends on the scaling parameter $\alpha$. 
Making this dependence explicit we redefine these functions as
\begin{align}
    \xi_n^\I(\phi)\rightarrow\sums_k \alpha^{s_{nk}^\I}\zeta_{nk}^\I(\phi) \,,
\label{eq:alpha_zeta}
\end{align}
where the range of the sum depends on the number of terms with different $\alpha$-orders. Furthermore, we add a coupling parameter to the derivative term as follows
\begin{align}
    (1+X/M_p^4)^n\rightarrow(1+\alpha^{r_n^\I}X/M_p^4)^n \,.
\label{eq:alpha_X}
\end{align}
Before substituting these redefinitions into Eq.~\eqref{eq:DGi2} we shall simplify the notation. 
Each power of $X/M_p^4$ in Eq.~\eqref{eq:DGi2} has a pre-factor of the form
\begin{align}
    \sums_{n=m}^{N^\I}\sums_{k}(-2)^m {{n}\choose{m}}\alpha^{mr_n^\I+s_{nk}^\I}\zeta_{nk}^\I(\phi) \,. \label{eq:prefactor_double_sum}
\end{align}
First, one can define a new function $\zeta_{mnk}^\I(\phi)$ such that 
$\zeta_{mnk}^\I(\phi) \equiv   (-2)^m{{n}\choose{m}}\zeta_{nk}^\I(\phi) $.
We then combine the double sums over $n$ and $k$ into a single sum over one index $k$. 
Furthermore, some terms in the double sum of Eq.~\eqref{eq:prefactor_double_sum} may cancel and so we only sum over the non-vanishing terms.
Taking this all into account, we can rewrite Eq.~\eqref{eq:DGi2} as
\begin{align}
    G_i(\phi,X)=\sums_{m= 0}^{N^\I}\left(\frac{X}{M_p^4}\right)^m \sums_{k\in I^\I_m} \alpha^{p_{mk}^\I} \zeta_{mk}^\I(\phi) \,,\label{eq:Gi_after_alpha}
\end{align}
where $k\in I^\I_m$ runs over the non-vanishing terms and $p_{mk}^\I=mr_{k}^\I + s_{k}^\I$. 
Since the scaling parameters in Eqs.~\eqref{eq:alpha_zeta} and \eqref{eq:alpha_X} can be chosen arbitrarily
we are free to impose $p_{mk}^\I\geq 0$ without loss of generality.

Ultimately, our goal is to find the order of $\alpha$ for each term of the field equations \eqref{eq:MFE_horndeski} and \eqref{eq:SFE_horndeski}. 
For this purpose, we utilise the operator $\alpha[\cdot]$ defined in Ref.~\cite{McManus:2016kxu} where $\alpha[F]$ gives the set of all orders of $\alpha$ contained in the expression $F$. 
For example, applying this operator to the expression  $F=A\alpha^p+B\alpha^q$ gives $\alpha[F]=\{p,q\}$.
Before applying this operator 
to Eq.~\eqref{eq:Gi_after_alpha}
we need to expand the scalar field around its background value $\phi=\phi_0(1+\alpha^q\psi)$. 
After this expansion the kinetic term becomes $X\rightarrow -\phi_0^2\alpha^{2q}(\partial\psi)^2/2$. 
Assuming that the functions $\zeta_{mk}^\I(\phi)$ are entire, i.e.~they can be written as an infinite polynomial sum, we can utilise the Weierstrass factorisation theorem 
\cite{Conway_functions} to write them as
\begin{align}
   \zeta_{mk}^\I(\phi)=(\phi-\phi_0)^{\mu_{mk}^\I}f_{mk}^\I(\phi) \,,\label{eq:factorization}
\end{align}
where $\mu_{mk}^\I\in \mathbb{N}$ is the multiplicity of $\phi_0$ and $f_{mk}^\I(\phi)$ is such that $f_{mk}^\I(\phi_0)\neq 0$. 
With the $\zeta$-functions written in this form, we can extract their $\alpha$-dependence as
\begin{align}
    \zeta_{mk}^\I(\phi)\rightarrow \alpha^{q\mu_{mk}^\I}\;\phi_0^{\mu_{mk}^\I}f_{mk}^\I(\phi_0) \, .
\end{align}
The $\alpha$-order of each $G_{i}$ function is then given by 
\begin{align}
    \alpha[G_i]&=\alpha\left[\sums_{(m,k)\in I^\I}\left(\frac{X}{M_p^4}\right)^m \alpha^{p_{mk}^\I} \zeta_{mk}^\I(\phi)\right]\nonumber\\[0.2cm]
    &=\bigcups_{(m,k)\in I^\I}\left\{p_{mk}^\I+(2m+\mu_{mk}^\I)q\right\} \,, \label{eq:alpha_Gi}
\end{align}
where $I^\I=\{(m,k)\in \mathbb{N}^2\ \vert\  m=0,...,N^\I;\ k\in I^\I_m \}$. 
As derivatives of the $G_{i}$-functions also enter the field equations it is necessary to determine their $\alpha$-orders too. 
These are given by 
\begin{subequations}
\begin{align}
    \alpha[G_{iX}]&=\bigcups_{(m,k)\in I_X^\I}\{p_{mk}^\I+(2m-2+\mu_{mk}^\I)q\} \,,\\
    \alpha[G_{i\phi}]&=\bigcups_{(m,k)\in I^\I}\{p_{mk}^\I+(2m+\mu_{1,mk}^\I)q\} \,,\\
    \alpha[G_{i\phi\phi}]&=\bigcups_{(m,k)\in I^\I}\{p_{mk}^\I+(2m+\mu_{2,mk}^\I)q\} \,,
\end{align}\label{eq:alpha_of_G}%
\end{subequations}
where $I_X^\I=\{(m,k)\in \mathbb{N}^2\ \vert\  m=1,...,N^\I;\ k\in I^\I_m \}$, $\mu_{1,mk}^\I$ and $\mu_{2,mk}^\I$ are the multiplicities of $\phi_0$ for the functions ${\zeta'}_{mk}^\I(\phi)$ and ${\zeta''}_{mk}^\I(\phi)$ respectively, where a prime denotes a derivative with respect to the scalar field $\phi$. 
Note that in the case where $\mu_{mk}^\I\neq 0$ 
it follows that $\mu_{1,mk}^\I=\mu_{mk}^\I-1$ and similarly $\mu_{1,mk}^\I\neq 0\implies \mu_{2,mk}^\I=\mu_{1,mk}^\I-1$. 
However when $\mu_{mk}^\I=0$ then $\mu_{1,mk}^\I$ can take any values in $\mathbb{N}$. 
When we consider the screening limit for large field value screening mechanisms $\alpha\rightarrow 0$, 
there can be functions of the form $\zeta_{mk}^\I=(\phi-\phi_0)^{1/n}$ or $\zeta_{mk}^\I=(\phi-\phi_0)^{-n}$ where $n\in\mathbb{N}^*$. 
We therefore need to include non-integer and negative multiplicities for large field value screening limits %
and so in this case we allow $\mu_{mk}^\I\in \mathbb{R}$. 

To conclude,
we shall find the mapping 
which relates the $\zeta^\I_{mk}(\phi)$ functions to $\xi_n^\I(\phi)$ by comparing Eq.~\eqref{eq:DGi2} to Eq.~\eqref{eq:Gi_after_alpha}.
This is given by 
\begin{align}
    \sums_{k\in I_\mu^\I}\alpha^{p_{mk}^\I}\zeta^\I_{mk}(\phi)=\sums_{n=m}^{N^\I}(-2)^m\xi_n^\I(\phi){{n}\choose{m}} \, , \quad\quad \forall m=0,...,N^\I \,.
    \label{eq:cond_inverse_mapping}
\end{align}
Since the set $\{\xi_n^\I(\phi)\}_{n\leq 2}$ satisfies the relations of Table~\ref{table:xi_n} we can use the nonlinear freedom in reconstructed Horndeski theories for $n>2$ to choose suitable functions that verify the conditions \eqref{eq:cond_inverse_mapping}.

\subsection{Einstein gravity limit for Horndeski theories with $c_{T}=1$}\label{subsec:Einstein_limit_luminal}

We now develop a method to determine whether a reconstructed Horndeski theory of the subclass $G_{4X}=G_{5}=0$ possesses an Einstein gravity limit. 
For this purpose, we distinguish the conditions for derivative screening and large field value screening for the following two reasons.
Firstly, because the screened region is represented by taking the different limits $\alpha\rightarrow\infty$ and $\alpha\rightarrow 0$ respectively. 
Secondly, because the allowed values for the multiplicities $\mu_{mk}^\I$, $i=2,3$ are different, namely $\mu_{mk}^\I\in \mathbb{N}$ for derivative screening and $\mu_{mk}^\I\in\mathbb{R}$ for large field value screening.

\subsubsection{Method for derivative screening}\label{subsubsec:derivative_screening_test}

Following the scaling method of Ref.~\cite{McManus:2016kxu}, 
we will now formulate constraints on $q$ such that we recover GR in the screening limit $\alpha\rightarrow\infty$. 
Examining the metric field equations in Eq.~\eqref{eq:MFE_horndeski}, we can see that in order to recover the Einstein field equations it is necessary that $G_4(\phi)=\phi$ is constant and non-zero in the screened limit. 
After expanding the scalar field $\phi=\phi_0(1+\alpha^q\psi)$ we can immediately see that $G_4\rightarrow \phi_0$ if $q<0$, which we take as the initial constraint on $q$. 
Furthermore, in order to recover GR in the screened region,
the sum $\sum R^\I_\munu$ must also vanish from the metric field equation in this limit. 
Thus we need to choose a value of $q$ such that all the terms in $\sum R^\I_\munu$ are of negative $\alpha$-order. 
Let $\mathcal{M}$ be the set of all orders in $\alpha$ of $\sum R^\I_\munu$,
\begin{align}
    \mathcal{M}:=\alpha\left[\sums_{i=2}^4R^\I_\munu\right] \, .
    \label{eq:definition_M}
\end{align}
Using the relations in Appendix~\ref{app:field_equations} between $R_\munu^\I$ and the Horndeski functions, one can directly rewrite $\mathcal{M}$ as
\begin{align}
    \mathcal{M}:=&\ \alpha[\alpha^q]\cup\alpha[G_2]\cup\alpha[\alpha^{3q}G_{3X}]\cup\alpha[\alpha^{2q}G_{3\phi}] \, ,\\[0.3cm]
    =&\ \{q\}\bigcups_{(m,k)\in I^{(2)}}\left\{p_{mk}^{(2)}+(2m+\mu_{mk}^{(2)})q\right\} \bigcups_{(m,k)\in I^{(3)}_X}\left\{p_{mk}^{(3)}+(2m+1+\mu_{mk}^{(3)})q\right\}\nonumber\\
    &\ \ \quad\bigcups_{(m,k)\in I^{(3)}}\left\{p_{mk}^{(3)}+(2m+2+\mu_{1,mk}^{(3)})q\right\} \, .
    \label{eq:M}
\end{align}
Note that the factors $\alpha^{nq}$ arise from the different derivatives of $\phi$ that accompany the Horndeski functions in the metric field equations and we made use of the relation $\alpha[\alpha^{2q}G_{2X}]\subset \alpha[G_2]$. 
We further define the set $Q_\mathcal{M}$ to contain all possible values of $q$ such that $0\in\mathcal{M}$, i.e.~all $q$-values for which there is one term in the metric field equations that is independent of $\alpha$.
Making use of Eq.~\eqref{eq:alpha_of_G} we find the set $Q_\m$ to be
\begin{align}
    Q_\mathcal{M}=&\{0\} \bigcups_{(m,k)\in I^{(2)}}\left\{\frac{-p_{mk}^{(2)}}{2m+\mu_{mk}^{(2)}}\right\} \bigcups_{(m,k)\in I^{(3)}_X}\left\{\frac{-p_{mk}^{(3)}}{2m+1+\mu_{mk}^{(3)}}\right\}\nonumber\\
    &\ \ \quad\bigcups_{(m,k)\in I^{(3)}}\left\{\frac{-p_{mk}^{(3)}}{2m+2+\mu_{1,mk}^{(3)}}\right\} \,.\label{eq:QM}
\end{align}
In order to end up with the Einstein field equations in the screened limit we need to find a condition on $q$
such that all the elements of $\mathcal{M}$ are strictly negative. 
Since all denominators in $Q_\mathcal{M}$ are positive, this requirement is simply 
\begin{align}
    q<q_\mathcal{M}:=\min Q_\mathcal{M} \, . \label{eq:cond_qM}
\end{align}
Note that certain denominators in $Q_\mathcal{M}$ could be zero and therefore require a more careful treatment.  
First note that zero denominators come from an element $c\in \m$ that is independent of $q$, implying there is a term in the metric field equations with the order $\alpha^c$. 
Due to the restriction that $p_{mk}^\I
\geq 0$ we have that $c\geq 0$. 
There are two scenarios to consider. 
When $c=0$ there is an undesired term in the metric field equation of order $\alpha^0$ and therefore we do not recover GR in the screened region. 
If on the other hand $c>0$ there is a term that diverges in the screened region and we cannot recover GR.
Note that in order for Eqs.~\eqref{eq:M} and \eqref{eq:QM} to hold we have assumed that $G_{2X}\neq 2G_{3\phi}$, a condition which is usually satisfied.
When dealing with a theory in which $G_{2X}=2G_{3\phi}$ one must remove from $\m$ and $Q_\m$ the contribution of $G_{3\phi}$ from the metric field equation as well as the term linear in $X$ from $G_{2}$.   

Let us now examine the scalar field equation \eqref{eq:SFE_horndeski}. 
We first note that for $G_4(\phi)=\phi$ the right-hand side of the scalar field equation is independent of $\alpha$. 
At least one term on the left-hand side that is also independent of $\alpha$ is therefore needed with every other term disappearing in the limit $\alpha\rightarrow\infty$. 
Furthermore, from the condition in Eq.~\eqref{eq:cond_qM} we know that $G_4\rightarrow\phi_0$ and $\sum R^\I\rightarrow 0$ in the screening limit. 
Consequently, we are left only with the terms multiplying $G_4$ on the left-hand side of the scalar field equation \eqref{eq:SFE_horndeski} to compensate the right-hand side. 
Let
\begin{align}
    \s:=\alpha\left[\sums_{i=2}^4\left(\nabla^\mu J_\mu^\I-P_\phi^\I\right)\right] \,,
\end{align}
and $Q_\s$ be the set of all possible values for $q$ such that $0\in \s$.
Our goal is to find the minimum value in $Q_\s$, which will subsequently be our value for $q$. 
In the same manner as for $\m$ one can write the set $\s$ as
\begin{align}
    \s=&\alpha[G_{2\phi}]\cup  \alpha[\alpha^{q}G_{2X}]\cup  \alpha[\alpha^{2q}G_{3\phi\phi}]\cup \alpha[\alpha^{2q}G_{3X}]
    \cup  \alpha[\alpha^{q}G_{3\phi}] \,,
\end{align}
where we again used the fact that $\alpha[\alpha^{2q}F_X]\subset\alpha[F]$ when $F$ can be written in terms of the expansion~\eqref{eq:Gi_after_alpha}. 
Here again note that in the case where $G_{2X}=2G_{3\phi}$, some terms in the scalar field equation cancel out, thus we need to remove $\alpha[\alpha^qG_{2X}]$ and for $G_{2X\phi}=G_{3\phi\phi}$, we need to remove $\alpha[\alpha^{2q}G_{3\phi\phi}]$. 
Keeping all the terms in the scalar field equation and using the results of Sec.~\ref{subsec:Exp_alpha_Gi} we find
\begin{align}
    Q_\s=&\bigcups_{(m,k)\in I^{(2)}}\left\{\frac{-p_{mk}^{(2)}}{2m+\mu_{1,mk}^{(2)}}\right\} \bigcups_{(m,k)\in I^{(2)}_X}\left\{\frac{-p_{mk}^{(2)}}{2m-1+\mu_{mk}^{(2)}}\right\}\bigcups_{(m,k)\in I^{(3)}}\left\{\frac{-p_{mk}^{(3)}}{2m+2+\mu_{2,mk}^{(3)}}\right\}\nonumber\\
    & \bigcups_{(m,k)\in I^{(3)}_X}\left\{\frac{-p_{mk}^{(3)}}{2m+\mu_{mk}^{(3)}}\right\} \bigcups_{(m,k)\in I^{(3)}}\left\{\frac{-p_{mk}^{(3)}}{2m+1+\mu_{1,mk}^{(3)}}\right\} \,. \label{eq:QS}
\end{align}
Since the aim is now to have a surviving term in the scalar field equation the requirement for $q$ becomes
\begin{align}
    q=q_\s:=\min Q_\s \,. \label{eq:cond_QS}
\end{align}
As was the case for $Q_\m$, $Q_\s$ can also contain elements with a zero in the denominator which we again separate into two distinct cases. 
The first case corresponds to terms that scale as $\alpha^{c}$ for some constant $c>0$, which diverge in the $\alpha \rightarrow \infty$ limit leading to an inconsistent theory. 
However, since we want to keep terms independent of $\alpha$, if $c=0$ we can still recover a consistent scalar field equation in the screened limit in contrast to the analogous case for the metric field equation.   
In this latter case, since one term in the scalar field equation already compensates the trace of the stress-energy tensor, the condition in Eq.~\eqref{eq:cond_QS} can be relaxed and we only require $q\leq q_\s$. 
We summarise our result with the following proposition: 
\begin{proposition}[Derivative screening]
A reconstructed Horndeski theory with $c_T=1$ has an Einstein gravity limit for $\alpha\rightarrow\infty$ $\iff$ $-\infty<q<q_\m$.
\end{proposition}\label{prop:derivative_screening}

\noindent Before considering large field value screening we shall make the method more efficient by reducing the number of relevant elements in $Q_\s$.
We initially note that either $\mu_{2,mk}^\3=\mu_{1,mk}^\3-1$ or $\mu_{2,mk}^\3\geq \mu_{1,mk}^\3$, therefore 
since the denominators are strictly positive we have that
\begin{align}
    \min\limits_{(m,k)\in I^{(3)}}\left\{\frac{-p_{mk}^{(3)}}{2m+1+\mu_{1,mk}^{(3)}}\right\}\leq\min\limits_{(m,k)\in I^{(3)}}\left\{\frac{-p_{mk}^{(3)}}{2m+2+\mu_{2,mk}^{(3)}}\right\} \,.
\end{align}
In addition, for $m\geq 1$ we can use a similar argument to show that %
\begin{align}
    \min\limits_{(m,k)\in I_X^{(2)}}\left\{\frac{-p_{mk}^{(2)}}{2m+\mu_{1,mk}^{(2)}}\right\}\geq \min\limits_{(m,k)\in I^{(2)}_X}\left\{\frac{-p_{mk}^{(2)}}{2m-1+\mu_{mk}^{(2)}}\right\} \,.
\end{align}
The term in $G_2(\phi,X)$ of order $m=0$ represents a scalar field potential, which can be shifted by a constant value with no physical implications. 
This implies that 
when $\mu_{0k}^\2=0$ and $\mu_{1,0k}^\2>0$ we can shift $G_2(\phi)$ such that $\mu^\2_{0k}=\mu_{1,0k}^\2+1$. 
Then the case of $m=\mu_{0k}^\2=0$ renders the metric field equations incompatible with GR and can be discarded. 
With these relations we can construct the set $\bar{Q}_\s\subset Q_\s$ such that $q_\s=\min\bar{Q}_\s$, i.e.~the smallest element of $Q_\s$ is also in $\bar{Q}_\s$. 
This smaller set is 
\begin{align}
    \bar{Q}_\s=\bigcups_{(m,k)\in I^{(2)}}\left\{\frac{-p_{mk}^{(2)}}{2m-1+\mu_{mk}^{(2)}}\right\} \bigcups_{(m,k)\in I^{(3)}_X}\left\{\frac{-p_{mk}^{(3)}}{2m+\mu_{mk}^{(3)}}\right\} \bigcups_{(m,k)\in I^{(3)}}\left\{\frac{-p_{mk}^{(3)}}{2m+1+\mu_{1,mk}^{(3)}}\right\} \,. \label{eq:QbarS}
\end{align}
Comparing $\bar{Q}_\s$ with $Q_\m$ in Eq.~\eqref{eq:QM} we can see that if $a/b\in Q_\m$, then $a/(b-1)\in \bar{Q}_\s$. 
We stress that the fraction $a/b$ is in general not irreducible and should not be reduced. For example, if $2/2\in Q_\m$, then $2/(2-1)\in Q_\s$ and not $1/(1-1)$.
In particular, this means that if $b=1$ then it must be the case that $a=0$ to avoid a divergent scalar field equation. 
Since we can always choose $q=q_\s$ even in the case where the scalar field equation contains an $\alpha^0$ term, this means that we only need $q_\m$ to find $q$. 
Indeed, if $q_\m=a/b$, then we can directly conclude that $q_\s=a/(b-1)=q$.

\subsubsection{Method for large field value screening}\label{subsubsec:large_field_value_test}

%
We shall now examine large field value screening mechanisms by considering the limit $\alpha\rightarrow 0$.
A key difference with derivative screening is that we allow the multiplicities to be non-integer and negative. 
This will have an implication on the method to test for a GR limit.
In particular, 
this results in some elements of $Q_\m$ and $Q_\s$ having negative denominators 
which will impact the inequality conditions.

As the method of obtaining $Q_\m$ and $Q_\s$ is independent of the particular screening limit we keep Eqs.~\eqref{eq:QM} and \eqref{eq:QS} as their respective definitions. 
As for derivative screening, one can find zero denominators in the elements of $Q_\m$ or $Q_\s$. 
By similar reasoning, we conclude that when $-\infty\in Q_\m$ or $-\infty\in Q_\s$ we still recover GR in the screened region. 
However, when there is a term independent of $\alpha$ in the metric field equations we do not recover the Einstein field equations. 
When there is a term independent of $\alpha$ in the scalar field equation we can still recover GR in the screened region and the condition $q=q_\s$ can also be relaxed. 
In order to differentiate between terms with negative or positive denominators we perform the following split
\begin{align}
    Q_\m&=Q_\m^-\cup Q_\m^+ \, ,\\
    Q_\s&=Q_\s^-\cup Q_\s^+ \, ,
\end{align}
where $Q_u^-$, $u=\m,\s$ contains elements with a denominator $\geq 0$ and $Q_u^+$ elements with a denominator $< 0$. 
The $\pm$ convention on $Q_u^\pm$ has been chosen so that elements of $Q_u^+$ are positive and elements of $Q_u^-$ are negative since the numerators are negative. 
As $Q_\m^-$ only contains zero or negative terms we have $\max Q_\m^-=0$. 
Furthermore, in order to only have strictly positive elements in $\m$ after the choice of $q$ we require
\begin{align}
    \max Q_\m^-=0<q<\min Q_\m^+:=q_\m^+ \,. \label{eq:chameleon_qm_condition}
\end{align}
As with derivative screening
we want to find $q_\s$ such that some terms in $\s$ are of order $\alpha^0$ while the $\alpha$-order of the other terms is greater than zero. 
This means that
\begin{align}
    q=q_\s^-:=\max Q_\s^-\qquad \textrm{or} \qquad q=q_\s^+:=\min Q_\s^+ \,.
\end{align}
However, since $Q_\s^-$ only contains elements that are less than or equal to zero the choice $q=q_\s^-$ cannot satisfy the condition in Eq.~\eqref{eq:chameleon_qm_condition} so it is discarded. 
We are then left with the condition
\begin{align}
    q=q_\s^+ \,,
\end{align}
which is relaxed to $q\leq q_\s^+$ if $\frac{0}{0}\in Q_\s$.
We summarise with the proposition:
\begin{proposition}[large field value screening]\label{prop:chameleon_screening}
A reconstructed Horndeski theory with $c_T=1$ has an Einstein gravity limit for $\alpha\rightarrow 0$ $\iff$ $0<q<q_\m^+$.
\end{proposition}
We can also simplify the method by analysing the set $Q_\s^+$. 
For any elements in $Q_\s^+$ we need $\mu_{mk}^\2$, $\mu_{1,mk}^\2$, or $\mu_{2,mk}^\2$ to be strictly smaller than zero, otherwise the denominator cannot be negative. 
Thus it is always the case that $\mu_{1,mk}^\2=\mu_{mk}^\2-1$ and $\mu_{2,mk}^\2=\mu_{1,mk}^\2-1$. 
This implies that we can find $q_\s^+=\min \bar{Q}_\s^+$, where 
\begin{align}
    \bar{Q}_\s^+\subset \bar{Q}_\s
\end{align}
only contains positive elements of $\bar{Q}_\s$. 
In contrast to derivative screening, it is not generally true that $q_\m^+:=a/b\implies q_\s^+=a/(b-1)$ because it is possible that there exists a negative element $c/d\in Q_\m^-$ such that $c/(d-1)\in \bar{Q}_\s^+$ is both positive and the smallest element of $\bar{Q}_\s^+$. 
In this case, one must construct $\bar{Q}_\s^+$ and find $q_\s^+=\min\bar{Q}_\s^+$.

\subsection{Application examples in reconstructed Horndeski theories}\label{subsec:examples}

To illustrate our method, 
we shall now apply it to reconstructed theories incorporating an example for each of the three classes of screening mechanisms, namely the chameleon, k-mouflage and Vainshtein mechanisms. 
As shown in Ref.~\cite{Kennedy:2019nie} we can implement these screening mechanisms through the nonlinear correction terms in reconstructed Horndeski theories. 
This means that it is always possible to choose a reconstructed theory which satisfies cosmological constraints while also exhibiting a screening mechanism. 
Such a mechanism can potentially make the reconstructed theory
compatible with Solar System experiments depending on the strength of the screening effect.

\subsubsection{Large field value screening: chameleon mechanism}\label{subsubsec:exple_chameleon}

We start by considering large field value screening via the chameleon mechanism, 
following Ref.~\cite{Kennedy:2019nie}. 
We only add a correction term to the Horndeski function $G_2(\phi,X)$ so that $G_3(\phi,X)=-2b_1(\phi)X$ and we set $G_4(\phi)=\phi$. 
With a suitable correction term we can replace the reconstructed potential $M_p^2U(\phi)$ with a chameleon potential of the form $\alpha^{N}(\phi-\phi_0)^k$. 
For example, we could have $G_{2}(\phi, X)$ of the form 
\begin{align}
    G_2(\phi,X)=-M_p^2U(\phi)+M_p^2Z(\phi)X+4a_2(\phi)X^2+\xi(\phi)\left(1-2\frac{X}{M_p^4}\right)^3 \,,\label{eq:G2_example}
\end{align}
where $\xi(\phi)=M_p^2U(\phi)-\alpha^{N}(\phi-\phi_0)^k$ and $N,k\geq 0$. 
We further assume that the functions $U(\phi)$, $Z(\phi)$ and $a_2(\phi)$ do not have an $\alpha$-dependence and they and their derivatives do not cancel at $\phi_0$. 
Without this assumption there may be another screening mechanism which conflicts with the chameleon mechanism we aim to implement.
To begin we write Eq.~\eqref{eq:G2_example} as an expansion in $X/M_p^4$,
\begin{align}
    G_2(\phi,X)=&\ \alpha^{N}(\phi-\phi_0)^k+\left\{-6M_p^6U(\phi)+6\alpha^{N}(\phi-\phi_0)^k+M_p^6Z(\phi)\right\}\frac{X}{M_p^4}+\nonumber\\
    &+\left\{12M_p^{10}U(\phi)-12\alpha^{N}(\phi-\phi_0)^k+4M_p^8a_2(\phi)\right\}\frac{X^2}{M_p^8}-8\xi(\phi)\frac{X^3}{M_p^{12}} \,.\label{eq:G2_example_2}
\end{align}
Examining the first term on the right-hand side of Eq.~\eqref{eq:G2_example_2} we see that $p_{00}^{(2)}=N$, $\mu_{00}^{(2)}=k$ and $\mu_{1,00}^{(2)}=k-1$. 
Similarly, from the second term we have $p_{1i}^{(2)}=\{0,N,0\}$, $\mu_{1i}^{(2)}=\{0,k,0\}$ and $\mu_{1,1i}^{(2)}=\{0,k-1,0\}$. 
We can proceed equivalently for the higher order terms. 
Using Eqs.~\eqref{eq:QM} and \eqref{eq:QbarS} the sets $Q_\m$ and $\bar{Q}_\s$ are given by
\begin{align}
    Q_\m&=\left\{0,\frac{-N}{k},\frac{-N}{2+k},\frac{-N}{4+k}\right\}\quad\textrm{and}\quad
    \bar{Q}_\s=\left\{0,\frac{-N}{k-1},\frac{-N}{1+k},\frac{-N}{3+k}\right\} \,.
\end{align}
Since we are dealing with large field value screening we follow the procedure derived in Sec.~\ref{subsubsec:large_field_value_test} and check for an Einstein gravity limit with proposition~\ref{prop:chameleon_screening}. 
We distinguish between separate cases: (i) when $N=0$ we have $Q_\m=Q_\s=\{0\}$ implying proposition \ref{prop:chameleon_screening} cannot be satisfied and conclude that $N>0$ in order to obtain an Einstein gravity limit; 
(ii) when $k=0$ there is a term $1\cdot\alpha^N$ in the action which cannot serve as a screening term since it gives no contribution to $Q_\s$, and $\bar{Q}_\s=\emptyset$. There is therefore no Einstein gravity limit; 
(iii) similarly for $k\geq 1$, we have $Q_\s^+=\emptyset$ implying there is no suitable choice of $q$;
(iv) with $0<k<1$ we have $q=q_\s^+=-N/(k-1)$ and $Q_\m^+=\emptyset$, which satisfy proposition \ref{prop:chameleon_screening}. 
We conclude that there only exists a GR limit when $N>0$ and $0<k<1$.

\subsubsection{First derivative screening: k-mouflage}\label{subsubsec:example_kmouflage}

To implement a screening mechanism via first derivatives of the scalar field we set the coupling strength parameter $\alpha=\alpha_K$. 
Referring to Appendix~\ref{app:reconstruction} we find the $\alpha$-dependence of the EFT functions to be $U,a_2,Z\sim \alpha^1$ and $b_1\sim\alpha^0$. 
Furthermore, we also assume here that these functions and their derivatives do not vanish at $\phi_0$. 
For clarity we extract the $\alpha$-dependence from the reconstructed functions so that, for example $U(\phi)=\alpha\, \bar{U}(\phi)$, to obtain the following Horndeski functions: $G_4=\phi$ and $G_3=-2\alpha^0 b_1(\phi)X$. 
For $G_2(\phi,X)$ we use the freedom to add a correction term of the form
\begin{align}
    \Delta G_2=\alpha M_p^2\bar{U}(\phi)\bigg(1+\frac{X}{M_p^4}\bigg)^3 \,,
\end{align}
where the $\alpha$-dependence was made explicit. 
This correction term has been chosen to cancel the potential in $G_2(\phi,X)$, 
avoiding a divergence in the limit $\alpha\rightarrow\infty$ because of its dependence on $\alpha$.
We can now expand the function $G_2$ and we obtain
\begin{align}
    G_2(\phi,X)=&\alpha\big(-6M_p^2\bar{U}(\phi)+M_p^6\bar{Z}(\phi)\big)\frac{X}{M_p^4}\nonumber\\&+\alpha\big(4M_p^8\bar{a}_2(\phi)+6M_p^2\bar{U}(\phi)\big)\bigg(\frac{X}{M_p^4}\bigg)^2-8\alpha M_p^2\bar{U}(\phi)\bigg(\frac{X}{M_p^4}\bigg)^3 \,.
\end{align}
Using the definitions in Eq.~\eqref{eq:QM} and Eq.~\eqref{eq:QS} we find the sets
\begin{align}
    Q_\m&=\bigg\{0,\frac{-1}{2},\frac{-1}{4},\frac{-1}{6}\bigg\} \quad \textrm{and}\quad
    \bar{Q}_\s=\bigg\{0,\frac{-1}{1},\frac{-1}{3},\frac{-1}{5}\bigg\} \, .
\end{align}
Following the procedure described in Sec. \ref{subsubsec:derivative_screening_test}, we have $q=q_\s=-1/1$ from $q_\m=-1/2$ and we recover GR in the screened region.

\subsubsection{Second derivative screening: Vainshtein mechanism}\label{subsubsec:example_vainshtein}

Finally, we implement a screening mechanism via second derivatives by taking the coupling parameter to be $\alpha=\alpha_{B}$~\cite{Kennedy:2019nie}.
Referring to Appendix~\ref{app:reconstruction} the functions in the reconstructed Horndeski theory are found to be of order $U,a_2,Z,b_1\sim \alpha^1$.
Furthermore, we also assume here that these functions and their derivatives do not cancel at $\phi_0$. 
Once again, we extract the $\alpha$-dependence of the EFT functions and we get the Horndeski functions: $G_4=\phi$ and $G_3=\alpha \bar{b}_1(\phi)X$. 
We cancel the leading and linear orders in $X$ of $G_2$ with the correction term 
\begin{align}
    \Delta G_2=&-\alpha M_p^6\bar{Z}(\phi) \left\{\big(1-2\frac{X}{M_p^4}\big)^4-\big(1-2\frac{X}{M_p^4}\big)^3\right\}\nonumber\\&+\alpha M_p^2\bar{U}(\phi)\left\{2\big(1-2\frac{X}{M_p^4}\big)^3-\big(1-2\frac{X}{M_p^4}\big)^6\right\} \,.
\end{align}
Once expanded, the function $G_2$ is given by
\begin{align}
    G_2(\phi,X)&=\alpha\big(4M_p^8\bar{a}_2-12M_p^6\bar{Z}-36M_p^2\bar{U}\big)\bigg(\frac{X}{M_p^4}\bigg)^2+\alpha\big(24M_p^6\bar{Z}+144M_p^2\bar{U}\big)\bigg(\frac{X}{M_p^4}\bigg)^3\nonumber\\
    &+\alpha\big(-16M_p^6\bar{Z}-240M_p^2\bar{U}\big)\bigg(\frac{X}{M_p^4}\bigg)^4+192\alpha M_p^2\bar{U}\bigg(\frac{X}{M_p^4}\bigg)^5-64\alpha M_p^2\bar{U}\bigg(\frac{X}{M_p^4}\bigg)^6 \,.
\end{align}
From this expression we see that $p_{mk}^\I=1$ and $\mu_{mk}^\I=\mu_{1,mk}^\I=\mu_{2,mk}^\I=0$, $\forall\ m,k,i$.
Using the definition in Eq.~\eqref{eq:QM} 
we find the set
\begin{align}
    Q_\m&=\bigg\{0,\frac{-1}{4},\frac{-1}{6},\frac{-1}{8},\frac{-1}{10},\frac{-1}{12},\frac{-1}{3}\bigg\} \,,
\end{align}
where the element $-1/3$ comes from $G_3(\phi,X)$. 
Following the procedure described in Sec.~\ref{subsubsec:derivative_screening_test} for derivative screening, we have $q=q_\s=-1/2$ from $q_\m=-1/3$ and we therefore recover GR in the screened region.

\section{Connecting PPN and EFT via reconstructed Horndeski models} \label{sec:PPN}

We have now shown how to implement various screening mechanisms in Horndeski theories reconstructed from the cosmological EFT of dark energy at lowest order in the scaling parameter $\alpha$.
This GR limit is required in order to satisfy the numerous astrophysical 
constraints from Solar System tests and pulsar timing observations \cite{Will:2014kxa,Renevey:2019jrm}. 
However, in order to describe corrections to the Einstein field equations we need to derive the metric field equations at next-to-leading order in $\alpha$.
We will now address how this is done for reconstructed Horndeski theories with $c_T=1$. 

In order to exploit the wealth of tests of GR on astrophysical scales
we wish to connect our framework to the PPN formalism.
For this purpose, we can use the corrected metric field equations to derive the corresponding set of PPN parameters of the reconstructed Horndeski models. 
To this end, we follow the method of Ref.~\cite{McManus:2017itv} 
where the PN expansion was derived for cubic Galileon and chameleon models. 
Note that we assume that the power of the scaling parameter $q$ has been found with the method in Sec.~\ref{sec:einstein_gravity_limit} and therefore each theory exhibits a screening mechanism.

We organise this section as follows. 
In Sec.~\ref{subsec:scaling_param_expansion} 
we expand the field equations in terms of the scaling parameter $\alpha$.
In Sec.~\ref{subsec:PN_expansion} we then derive the PN expansion of the metric field equations. 
Finally, in Sec.~\ref{subsec:PPN_param} we compare our PN expansion to the PPN formalism of Ref.~\cite{Will:2018bme} to derive the PPN parameters and the effective gravitational constant for reconstructed Horndeski theories. 
The provides a connection between the cosmological EFT of dark energy and modified gravity and the PPN formalism within the framework of Horndeski gravity.

\subsection{Expansion in the scaling parameter}\label{subsec:scaling_param_expansion}

We shall first provide the lowest-order screening correction in Sec.~\ref{subsubsec:lowest_order} and then the next-to-leading order correction in Sec.~\ref{subsubsec:higher_order}.

\subsubsection{Lowest-order screening correction}\label{subsubsec:lowest_order}

Recall that in order to find the field equations in the screened region the first step is to expand the scalar field as $\phi\rightarrow \phi_0(1+\alpha^q\psi)$.
Since we only consider theories which reduce to GR in the screening limit, 
at order $\alpha^0$ the metric field equations must correspond to GR,
\begin{align}
    \phi_0R_\munu^\0=8\pi G\left(T_\munu^\0-\frac{1}{2}T^\0g_\munu^\0\right) \, ,\label{eq:metric_eqt_lowest_order}
\end{align}
where the index $\0$ represents the screening order of the metric $g_\munu\rightarrow \alpha^0 g_\munu^\0$. 

It is not as straightforward to find the scalar field equation at lowest order.
In contrast to the metric field equations, the leading order in the scalar field equation is $\alpha^q$ with several terms surviving in the screening limit. 
These surviving terms 
depend on the particular screening mechanism and a priori there is no method to identify them in general.
However, there are still few simplifications we can make. 
Examining Eq.~\eqref{eq:SFE_horndeski} we know from the restrictions for the metric field equations that $G_4\rightarrow \phi_0$ and $\sum R^\I\rightarrow 0$.
Therefore we only need consider the term multiplying $G_4$ on the left-hand side of Eq.~\eqref{eq:SFE_horndeski}. 
Since we cannot guess which terms will survive in the screening limit we include a delta-function $\delta:\mathbb{R}\rightarrow\{0,1\}$, such that $\delta(0)=1$ and zero otherwise, in front of each term. 
These $\delta$-functions will vanish when the power of $\alpha$ differs from the required order $\alpha^q$. 
In this manner, a general expression for the scalar field equation at lowest order in the screening correction can be written as
\begin{align}
   \frac{-8\pi GT}{\phi_0}&=\sums_{(m,k)}\left(2G^\3_{mk\phi}\delta((2m+1+\mu_{1,mk}^\3)q+{p}_{mk}^\3)-G^\2_{mkX}\delta((2m-1+\mu_{mk}^\2)q+{p}_{mk}^\2)
   \right)\Box\phi\nonumber\\
   &+\sums_{(m,k)}\left(G^\2_{mk\phi X}\delta((2m+\mu_{1,mk}^\2)q+{p}_{mk}^\2)-G^\3_{mk\phi\phi}\delta((2m+2+\mu_{2,mk}^\3)q+{p}_{mk}^\3)
   \right)2X\nonumber\\
   &+\sums_{(m,k)}\left(2G^\3_{mk\phi X}\delta((2m+1+\mu_{1,mk}^\3)q+{p}_{mk}^\3)-G^\2_{mkX X}\delta((2m-1+\mu_{mk}^\2)q+{p}_{mk}^\2)
   \right) \nonumber\\
   &\times \nabla_\mu X\nabla^\mu \phi -\sums_{(m,k)}G^\2_{mk\phi}\delta((2m+\mu_{1,mk}^\2)q+{p}_{mk}^\2)
   \nonumber\\
   &-\sums_{(m,k)}2G^\3_{mk\phi X}\delta((2m+1+\mu_{1,mk}^\3)q+{p}_{mk}^\3)X\Box\phi
   \nonumber\\
   &+\sums_{(m,k)}G^\3_{mkXX}\delta((2m+\mu_{mk}^\3)q+{p}_{mk}^\3)\left(\Box\phi\nabla_\mu\phi\nabla^\mu X+\nabla_\mu X\nabla^\mu X\right)
   \nonumber\\
   &+\sums_{(m,k)}G^\3_{mkX}\delta((2m+\mu_{mk}^\3)q+{p}_{mk}^\3)\left((\Box\phi)^2-\nabla_\mu\nabla_\nu\phi\nabla^\mu\nabla^\nu\phi\right) \,, \label{eq:scalar_eqt_lowest_order}
\end{align}
where we kept implicit the screening order index $\0$ of the metric for simplicity and each sum runs over $I^\J$ (see Sec.~\ref{subsec:Exp_alpha_Gi}). 
We also defined
\begin{align}
    G^\J_{mk}=G^\J_{mk}(\phi,X):=\left(\frac{X}{M_p^4}\right)^m \zeta_{mk}^\J(\phi)
\end{align}
to simplify the notation. 
Written this way Eq.~\eqref{eq:scalar_eqt_lowest_order} 
appears rather cumbersome, though most of the $delta$-functions will vanish for a particular theory. 
Eq.~\eqref{eq:scalar_eqt_lowest_order} governs the behaviour of the scalar field perturbation in the screened region at order $\alpha^q$. 
Our goal is now to find the next-to-leading order metric field equations describing the correction terms to the metric.

\subsubsection{Next-to-leading order screening correction of the metric field equations}\label{subsubsec:higher_order}

We will eventually demonstrate that the order of the field equations corresponding to first order corrections to the metric is $\alpha^q$. 
For now, we expand the metric in terms of some as yet unknown order $\alpha^{p}$
\begin{align}
    g_\munu=g_\munu^\0+\alpha^p g_\munu^\p,\label{eq:metric_expansion}
\end{align}
where $g_\munu^\0$ is the background field satisfying the Einstein field equations and $\alpha^p g_\munu^{(p)}$ is the correction term. 

Our ultimate aim is to derive the field equations governing the behaviour of $g_\munu^\p$.
To this end one must initially determine the power $p$ of the $\alpha$-order of the correction term.
Here we demonstrate how to derive $p$ in the case of derivative screening with $\alpha \rightarrow \infty$, 
though note that by interchanging $\min\{\}\leftrightarrow \max\{\}$ and the inequalities $<\ \leftrightarrow\ >$ the following derivation can be applied to large field value screening with $\alpha\rightarrow 0$.
Taking the metric field equations~\eqref{eq:MFE_horndeski} we expand the scalar field and the metric as $\phi\rightarrow\phi_0(1+\alpha^q\psi)$ and $g_\munu\rightarrow g_\munu^\0+\alpha^pg_\munu^\p$ respectively. 
The terms generated by the background metric $g_\munu^\0$ are of order $\alpha^0$ and $\alpha^k$, where $k\in\m$ and $\m$ are defined in Eq.~\eqref{eq:definition_M}. 
Hence the terms generated by $g_\munu^\p$ are of order $\alpha^p$ and $\alpha^{p+k}$. 
As the elements of $\m$ are negative, the slowest converging power in $\m$ in the limit $\alpha\rightarrow\infty$ is $\max\m$. 
In addition, there is only one term that is strictly of order $\alpha^p$, namely $\alpha^p\phi_0R_\munu^\p$.
Therefore $p$ must correspond to $\max\m$ to ensure the field equations are consistent in the screening limit so that $p=\max\m$.
Let's now find the relation between $\max\m$ and $q$. 
Recall the elements of $\m$ take the form $a_l+b_lq$ where $l$ runs over the elements of $\m$ and $q$ is the smallest element of $Q_\s$. 
From the definition of $Q_\m$ in Eq.~\eqref{eq:QM} and $Q_\s$ in Eq.~\eqref{eq:QS} we have that $-a_l/b_l\in Q_\m$ and $-a_l/(b_l-1)\in Q_\s$. 
If we define $q=-a/(b-1)$, we have that $-a_l/(b_l-1)\geq -a/(b-1)$, $\forall\, l$, since $q$ is the smallest element of $Q_\s$. 
Using these relations we determine that any element of $\m$ satisfies
\begin{align}
    a_l+b_lq=\frac{a_l(b_l-1)}{b_l-1}-\frac{b_la}{b-1}\leq\frac{a(b_l-1)}{b-1}-\frac{b_la}{b-1}=q \,,
\end{align}
and since $q\in \m$ we can conclude that
\begin{align}
    p=\max\m=q \,.\label{eq:p=q}
\end{align}
This result also holds for the screening limit $\alpha\rightarrow 0$. 
Even though Eq.~\eqref{eq:p=q} appears trivial, it will have implications when proceeding with the PN expansion. 
In particular, it implies the PPN parameter $\gamma$ and the effective gravitational constant will have the same form in any theory with a screening mechanism described by the scaling method.

We shall now write the metric field equations 
satisfied by the metric correction term $g_\munu^\q$. 
As with the scalar field equation we do not know a priori which terms in the metric field equations~\eqref{eq:MFE_horndeski} are of order $q$. 
Consequently, we also need to include $\delta$-functions. 
Since $m\in\m\implies m<0$, each term of the form $m+q$
is of a smaller order than $q$ and thus does not contribute. 
Additionally, any terms of the form $2q$, $3q$ and so on are also of a smaller order than $q$ and are discarded. 
One can then write the general metric field equations at order $q$ as
\begin{align}
    R_\munu^\q&=\frac{8\pi G\mathcal{T}_\munu^\q}{\phi_0}-\psi R_\munu^\0+\nabla_\mu^\0\nabla_\nu^\0\psi+\frac{1}{2}g_\munu^\0\Box^\0\psi\nonumber\\
    &+\frac{1}{2}\sums_{(m,k)}G^{(2)(0)}_{mkX}\delta((2m+\mu_{mk}^\2-1)q+p_{mk}^\2)\nabla_\mu^\0\phi\nabla_\nu^\0\psi\nonumber\\
    &+\frac{1}{2\phi_0}\sums_{(m,k)}(m-1)G^{(2)(0)}_{mk}\delta((2m+\mu_{mk}^\2-1)q+p_{mk}^\2)g^\0_\munu\nonumber\\
    &-\frac{1}{2}\sums_{(m,k)}G^{(3)(0)}_{mkX}\delta((2m+\mu_{mk}^\3)q+p_{mk}^\3)\times\nonumber\\[-1em]&\qquad\qquad\times\left(\Box^\0\phi\nabla^\0_\mu\phi\nabla^\0_\nu\psi+2\nabla^\0_{(\mu} X^\0\nabla^\0_{\nu)}\psi+g_\munu^\0 X^\0\Box^\0\psi\right)\nonumber\\[0.6em]
    &-\sums_{(m,k)}G^{(3)(0)}_{mk\phi}\delta((2m+1+\mu_{1,mk}^\3)q+p_{mk}^\3)\nabla^\0_\mu\phi\nabla^\0_\nu\psi \,.\label{eq:higher_order_metric_eqt}
\end{align}
As before, Eq.~\eqref{eq:higher_order_metric_eqt} simplifies for particular examples.
Note that we have also used the relation $\nabla_\mu\phi=\phi_0\nabla_\mu\psi$ to simplify the notation where possible.

Eq.~\eqref{eq:higher_order_metric_eqt} describes the behaviour of the correction term $g_\munu^\q$. 
We will need to expand this term to PN order in order to compare it with the PPN metric~\eqref{eq:PPN_metric}. 
Note that in the opposite limit $\alpha\rightarrow 0$ Eqs.~\eqref{eq:scalar_eqt_lowest_order} and \eqref{eq:higher_order_metric_eqt} are the same, where in this case $q>0$ and $q=\min\m$.

\subsection{Post-Newtonian expansion of the metric field equations}\label{subsec:PN_expansion}

Having obtained the metric field equations at order $\alpha^{0}$ and $\alpha^{q}$ in Eqs.~\eqref{eq:metric_eqt_lowest_order} and \eqref{eq:higher_order_metric_eqt} respectively, we shall now derive the PN solutions for $g_\munu^\0$ and $g_\munu^\q$. 
Note that the PN expansion 
is independent of whether the limit $\alpha \rightarrow 0$ or $\alpha\rightarrow \infty$ is taken.
Before commencing we shall define some useful notation. 
Since we are dealing with two distinct expansions in $\alpha$ and in $v\sim (GM/R)^{1/2}$, we denote the orders of these expansions as $\alpha^q\sim\oa(q)$ and $v^i\sim\opn(i)$ respectively. 
Moreover, the $\alpha$-order of the scalar perturbation $\psi$ is $\oa(q)$ and we write its PN order as $\psi^\I\sim\opn(i)$. 
Similarly, the order of the metric perturbation is written $g_\munu^\q=h_\munu^{(q,k)}\sim\opn(k)$.

\subsubsection{Lowest-order PN expansion}\label{subsubsec:lowest_PN}

As $g_\munu^\0$ is of order $\oa(0)$ and satisfies the Einstein field equations, we can directly use the result in Ref.~\cite{Will:2018bme} to find the PN expansion of the metric. 
At lowest order in the PN expansion, the scalar and metric perturbations are of the form $\psi=\psi^\I$ and $g_\munu^\q=h_\munu^{(q,k)}$, where $i$ and $k$ are a priori independent. 
Given that the following gauge fixing conditions of scalar-tensor theories \cite{McManus:2017itv}
\begin{subequations}
\begin{align}
    g_{i,\mu}^\mu-\frac{1}{2}g^\mu_{\mu,i}&=\psi_{,i}\, ,\\
    g_{0,\mu}^\mu-\frac{1}{2}g^\mu_{\mu,0}&=\psi_{,0}-\frac{1}{2}g_{00,0} \, ,
\end{align}\label{eq:gauge_conditions}%
\end{subequations}
need to be satisfied at each order in $\alpha$ and $v$ we conclude that $k=i\implies g_\munu^\q,\psi\sim\opn(i)$ at lowest order in $v$.  
We may also infer this result by examining the first line of the metric field equations~\eqref{eq:higher_order_metric_eqt}. 
To obtain the PN order $i$ of the scalar field perturbation $\psi^\I$ we first express Eq.~\eqref{eq:scalar_eqt_lowest_order} in terms of $\psi^\I$.  
Combining the requirement that the PN order must be equal on the left-hand side and right-hand side with the fact that the lowest PN order of each term is $T=-\rho\sim \opn(2)$~\cite{Will:2018bme}, $g_\munu^\0=\eta_\munu$ and $\psi=\psi^\I$, we can derive the PN order $i$.
In general this procedure is rather involved but substantially simplifies for specific examples such as that presented in Appendix~\ref{app:pulsar_test}.
Before continuing note that if the fourth line of Eq.~\eqref{eq:scalar_eqt_lowest_order} has a non-vanishing term with $X^m$ and $m=0$, then the right-hand side contains a constant term of PN order $\opn(0)$. 
Since $T$ is of order $\opn(2)$ this would lead to a contradiction.
However, this is only the case when the Horndeski action contains a potential term of the form $(\phi-\phi_0)$. 
As we have seen in the example of Sec.~\ref{subsubsec:exple_chameleon}, such a term alone cannot lead to screening but it would survive in the screening limit since it is proportional to $\alpha^q$. 
In what follows, we do not consider theories with such a potential term.
We assume the matter content to be a perfect fluid, 
which implies the stress-energy tensor at $\opn(4)$ is given by~\cite{McManus:2017itv}
\begin{align}
    T_{00}=&\rho(1+\Pi+v^2-h_{00}^{(0,2)}) \,,\\
    T_{0i}=&-\rho v^i \,,\\
    T_{ij}=&\rho v^iv^j+p\delta^{ij} \,.
\end{align}
Using the virial relation and the Poisson equation one finds that $\rho\sim\opn(2)$ and $p\sim\opn(4)$, where $\rho$ is the matter density of the fluid, $p$ is the pressure and $\Pi\sim\opn(2)$ is any energy density in the fluid other than $\rho$. 

We are now ready to proceed with the PN expansion of the perturbed metric field equations~\eqref{eq:higher_order_metric_eqt}. 
To begin, note that $R_\munu^\q=R_\munu^{(q,i)}$ implying we only retain terms of order $\opn(i)$ on the right-hand side of Eq.~\eqref{eq:higher_order_metric_eqt}. 
Additionally, at lowest PN order we have that $R_{00}^{(q,i)}=-\frac{1}{2}\nabla^2 h_{00}^{(q,i)}$ \cite{Straumann:2013spu} where $\nabla^2$ is the three-dimensional Laplacian operator. 
Thus we can write the 00-component of Eq.~\eqref{eq:higher_order_metric_eqt} at order $\opn(i)$ as
\begin{align}
    \nabla^2 h_{00}^{(q,i)}=\nabla^2 \psi^\I  \, . 
\label{eq:h00_order_i}
\end{align}
Eq.~\eqref{eq:h00_order_i} can be trivially solved to yield
\begin{align}
    h_{00}^{(q,i)}=\psi^\I \, .
\end{align}
We then proceed in a similar manner for $h_{ij}^{(q,i)}$ finding that
\begin{align}
    h_{ij}^{(q,i)}=-\psi^\I\delta_{ij} \, .
\label{eq:hij_order_i}
\end{align}
The simple form of Eq.~\eqref{eq:h00_order_i} comes from the result $p=q$ in Eq.~\eqref{eq:p=q}. 
In the absence of this equality there would be additional $\delta$-functions of the form $\delta(p-q)$.

There is a consistency issue in our PN formalism in that the PN order of the correction term $h_{00}^{(q,i)}\sim \opn(i)$ generally does not match the usual PN orders of the 00-component of the metric, namely $\opn(0)$, $\opn(2)$ and $\opn(4)$. 
However, the full correction term includes a contribution from the coupling parameter given by $\alpha^q h_{00}^{(q,i)}$. 
In order for the correction term to have an allowed PN order
the coupling parameter $\alpha$ must itself possess a PN order. 
Since the metric should be Minkowski at order $\opn(0)$, $\alpha$ should be of order $\opn\left((2-i)/q\right)$ for the correction term to be of order $\opn(2)$. 
This is a reasonable assumption as it was shown to be the case for Galileon interactions in Ref.~\cite{Bolis:2018kcq}.

\subsubsection{Higher-order PN expansion for the 00-component}\label{subsubsec:higher_PN}

Ultimately, our primary interest is to calculate the PPN parameters $\gamma$ and $\beta$ for which we only need the PN expansion of $g_{00}$ and $g_{ij}$.
We can therefore omit the calculation of the PN expansion of $g_{0i}$ and derive $g_{00}$ at higher order, noting that $g_{ij}$ is already at PN order.
At the next PN order $l$ the 00-component of the metric becomes $g_{00}^\q=h_{00}^{(q,i)}+h_{00}^{(q,l)}$ and we keep the scalar field at order $\opn(i)$ as in Ref.~\cite{Will:2018bme}. 
By analysing the metric field equations~\eqref{eq:higher_order_metric_eqt} we see that terms of the form $\nabla_\mu\nabla_\nu\psi$ or $\Box\psi$ give rise to expressions of order $\opn(i+2)$, such as $\partial_0^2\psi$, implying the next PN order is $l=i+2$. 
This is consistent with the usual PN expansion, since the PN order increases by two for each order in the PPN formalism. 
Now that we know the PN order of 
Eq.~\eqref{eq:higher_order_metric_eqt}
we perform the PN expansion,
keeping terms of order $\opn(2+i)$ on the right-hand side. 
This leads to
\begin{align}
    \nabla^2 h_{00}^{(q,i+2)}=&-\partial_0^2\psi-h_{00}^{(0,2)}\nabla^2\psi+\frac{16\pi G\rho}{\phi_0}h_{00}^{(q,i)}-\psi\nabla^2 h_{00}^{(0,2)}\nonumber\\[0.7em]
    &+\psi_{,j}h_{00,j}^{(0,2)}-2h_{00,j}^{(q,i)}h_{00,j}^{(0,2)}+h_{jk}^{(0,2)}h_{00,jk}^{(q,i)}+h_{jk}^{(q,i)}h_{00,jk}^{(0,2)}\nonumber\\[1em]
    &+\sums_{(m,k)\in I^\2} \Psi_{mk}^{(2)}\delta((2m+\mu_{mk}^\2-1)q+p_{mk}^\2)\delta((2m-1+\mu_{mk}^\2)i-2)\nonumber\\
    &-\sums_{(m,k)\in I^\3} \Psi_{mk}^{(3)}\nabla^2\psi\delta((2m+\mu_{mk}^\3)q+p_{mk}^\3)\delta((2m+\mu_{mk}^\3)i-2) \, ,\label{eq:h00_i+2}
\end{align}
where we defined 
\begin{align}
    \Psi_{mk}^{(2)}&=\frac{m-1}{2}\left(\frac{\phi_0\grad\psi}{M_p^2}\right)^{2m}\frac{f_{mk}^\2(\phi_0)}{\phi_0^{1-\mu_{mk}^\2}}{\psi^{\mu_{mk}^\2}} \, , \label{eq:Mmk2}\\ 
    \Psi_{mk}^{(3)}&=m\left(\frac{\phi_0\grad\psi}{M_p^2}\right)^{2m}f_{mk}^\3(\phi_0)\phi_0^{\mu_{mk}^\3}{\psi^{\mu_{mk}^\3}} \, ,
\label{eq:Mmk3}
\end{align}
and $f_{mk}^\J(\phi)$ is defined in Eq.~\eqref{eq:factorization}. 
We also used the gauge conditions in Eqs.~\eqref{eq:gauge_conditions} and the Ricci tensor at order $\opn(i+2)$ \cite{Straumann:2013spu} 
\begin{align}
    R_{00}^{(q,i+2)}=-\frac{1}{2}\bigg(\nabla^2 h_{00}^{(q,i+2)}-2\psi_{,00}-\psi_{,j}h_{00,j}^{(0,2)}+2h_{00,j}^{(q,i)}h_{00,j}^{(0,2)}-h_{ij}^{(0,2)}h_{00,jk}^{(q,i)}-h_{ij}^{(q,i)}h_{00,jk}^{(0,2)}\bigg) \, .
\end{align}
Eq.~\eqref{eq:h00_i+2} can again be solved using an inverse Laplacian yielding \cite{McManus:2017itv}
\begin{align}
   h_{00}^{(q,i+2)}&=\Tilde{\Phi}_1-3\mathcal{A}_\psi-\mathcal{B}_\psi+6\frac{G}{\phi_0}\Tilde{\Phi}_2+\Phi_{Sc} \nonumber \\
   &:=\Phi_\textrm{BD}+\Phi_\textrm{Sc} \, ,\label{eq:h00_higher_order}
\end{align}
where $\Phi_\textrm{Sc}$, the inverse Laplacian of the final two lines of Eq.~\eqref{eq:h00_i+2}, is called the screening potential and the inverse Laplacian of the other terms are collected in the potential $\Phi_\textrm{BD}$.
They are defined in Appendix~\ref{app:NP} and calculated in Ref.~\cite{McManus:2017itv}. 
For completeness we shall write the partial differential equation (PDE) which one solves for the screening potential.
In Eq.~\eqref{eq:h00_i+2}, we can see that the final two terms contain two $\delta$-functions each, one depending on the $\alpha$-order $q$ and the other one on the PN order $i$. 
In the case where a single term 
drives the screening mechanism such that $q$ only satisfies one $\delta$-function in Eq.~\eqref{eq:h00_i+2}, the $\delta$-function depending on the PN order $i$ will also be satisfied for this particular term. 
This comes from the fact that, for the surviving term with indices $(\bar{m},\bar{k})$, we can relate both orders $i$ and $q$ with the relation $i=-2q/p_{\bar{m},\bar{k}}$.
Since we will only consider cases where one term is responsible for screening
this leads to the PDE
\begin{align}
    \nabla^2\Phi_\textrm{Sc} = & \sums_{(m,k)\in I^\2} \Psi_{mk}^{(2)}\delta((2m+\mu_{mk}^\2-1)q+p_{mk}^\2) \nonumber\\
    & -\sums_{(m,k)\in I^\3} \Psi_{mk}^{(3)}\nabla^2\psi^\I\delta((2m+\mu_{mk}^\3)q+p_{mk}^\3) \,.\label{eq:PDE_Phi_Sc}
\end{align}
If more than one term contributes to the screening mechanism, but they all have the same parameter $p_{mk}$, Eq.~\eqref{eq:PDE_Phi_Sc} is also valid. 
However, if they do not have the same $p_{mk}$ one has to pay attention to the PN order $i$ and add the other $\delta$-function that depends on $i$.

\subsection{Mapping to the PPN parameters}\label{subsec:PPN_param}

Finally, we can now derive the PPN parameters for Horndeski theories reconstructed from the cosmological effective field theory of dark energy and modified gravity in their screened regime.
By virtue of the fact that the PPN parameters $\gamma$ and the effective gravitational constant $G_\eff$ are derived from the metric field equations at lowest PN order,
we begin by deriving their form from Eqs.~\eqref{eq:h00_order_i} and \eqref{eq:hij_order_i}.
With the usual PN expansion of GR \cite{Will:2018bme} up to $\opn(2)$ this gives 
\begin{align}
    g_{00}&\approx -1+\frac{2G}{\phi_0}\Phi_N+\alpha^q\psi \, ,\\
    g_{ij}&\approx \delta_{ij}\left(1+\frac{2G}{\phi_0}\Phi_N-\alpha^q\psi\right) \, ,
\end{align}
where we recall $\Phi_N=\Phi_N(\vb{x})$ is the Newtonian potential defined in Eq.~\eqref{eq:appendix_NP} sourced from $h_{00}^{(0,2)}$. 
Comparing our results to the PPN metric at $\opn(2)$
\cite{Will:2018bme}
\begin{align}
    g_{00}&\approx -1+2G_\eff \Phi_N \, ,\\
    g_{ij}&\approx \delta_{ij}(1+2\gamma G_\eff \Phi_N) \, ,
\end{align}
leads to the following solution for $G_\eff$ and $\gamma$
\begin{align}
    G_\eff&:=G^\0+G^\q=\frac{G}{\phi_0}+\frac{\alpha^q\psi}{2\Phi_N} \, ,\label{eq:Geff_PPN}\\
    \gamma&:=\gamma^\0+\gamma^\q=1-\frac{\alpha^q\psi}{G_\eff \Phi_N} \, .
\label{eq:gamma}
\end{align}
We now consider 
the PPN parameter $\beta$ for which it is necessary to examine the next order in the PN expansion.
Following the same method used to derive $G_\eff$ and $\gamma$, $\beta$ is obtained by comparing the metric solution in Eq.~\eqref{eq:h00_i+2} to the standard PPN metric.
We further normalise the value of the effective gravitational constant such that $G_\eff=1$. 
Using the PN expansion of GR 
we can express $g_{00}$ up to PN order $\opn(4)$ as
\begin{align}
    g_{00} \approx & -1+2\Phi_N-2{G^\0}^2\Phi_N^2+4G^\0\Phi_1-2{G^\0}^2\Phi_2+2G^\0\Phi_3+6G^\0\Phi_4 \nonumber\\
    & +\alpha^q\left(\Phi_{BD}+\Phi_{Sc}\right) \, ,
\label{eq:g00_i+2}
\end{align}
where the PN potentials $\Phi_i$ are defined in Eq.~\eqref{eq:appendix_NP}. 
Comparing our result to the standard PPN metric gives
\begin{align}
    g_{00}\approx&-1+2\Phi_N-2\beta \Phi_N^2+(2\gamma+2)\Phi_1\nonumber\\&+2(-2\beta+1)\Phi_2+2\Phi_3+6\gamma\Phi_4
    \, ,
\label{eq:g00_PPN4}
\end{align}
where the other PPN parameters are set to zero. 
A comparison between Eq.~\eqref{eq:g00_i+2} and Eq.~\eqref{eq:g00_PPN4} yields $\beta$ to be 
\begin{align}
    \beta := & \beta^\0+\beta^\q\nonumber\\
     = & 1+\alpha^q \left[4\Phi_N^2\left(2\Phi_2+\Phi_N^2\right)\right]^{-1} \left[
    \alpha^q{\psi}^2\left(\Phi_2+\Phi_N^2-6\Phi_4\right) \right. \nonumber\\
    & \left. -\Phi_N \psi\left(\Phi_1+4\Phi_2-2\Phi_3-6\Phi_4+4\Phi_N^2\right)-2\Phi_N^2\left(\Phi_{BD}+\Phi_{Sc}\right) \right] \, .
\label{eq:beta}
\end{align}
Note that in this framework the PPN parameters depend on the position and time of the objects subjected to a gravitational field. 
This is to be expected, as to implement a successful screening mechanism the scalar field must vary with distance or density. 
As a result, the PPN parameters must also vary as they are influenced by 
the scalar field. 
A further feature of this framework is its universality. 
Recall that in the development of the PPN formalism for reconstructed theories the results are 
independent of whether the limit $\psi\ll \alpha$ or $\psi\gg \alpha$ is taken. 
The only differences arise in the scalar field equation for $\psi$ and the form of the screening potential $\Phi_\textrm{Sc}$.

\section{Conclusions} \label{sec:conclusions}
%
Substantial and complementary efforts are being made to perform precise tests of GR on astrophysical and cosmological scales.
Due to the sheer size of the gravitational model space 
it is of great importance that generalised and efficient techniques are developed that enable effective discrimination between competing theories
with the observational data.
In order to avoid bias towards certain models, many generic parameterisations have been devised which enable one to study the physical implications of a large class of models without restricting to a particular theory. 
For example, in cosmology the EFT of dark energy and modified gravity is widely used to study the dynamics of the cosmological background and perturbations
in a generic manner.
On astrophysical scales, the PPN formalism provides
an effective tool for model-independent tests of GR.
It is indeed important to study the observational implications of the model space across a wide range of length scales.
This point was emphasised for example in Ref.~\cite{Kennedy:2019nie}, where it was demonstrated that
Horndeski scalar-tensor theories cannot exhaustively be distinguished from $\Lambda$CDM up to arbitrary finite $n$-th order in the cosmological perturbations.
Astrophysical constraints are therefore needed to complement cosmological constraints and vice-versa. 
Upcoming cosmological surveys such as Euclid and LSST will place new constraints on the EFT parameter space.
Through the reconstruction method of Refs.~\cite{Kennedy:2017sof, Kennedy:2018gtx} it will furthermore be possible to directly place general cosmological constraints on the form of the Horndeski functions themselves. 
Particularly strong and general bounds on deviations from GR have been placed on astrophysical scales with tight constraints on the PPN parameters.
For example, pulsar observations have recently proved to be particularly effective at imposing
bounds on the PPN parameters. 
New experiments, such as the FAST or SKA radio telescopes, are set to detect new pulsar system which will further tighten these constraints.  
New Solar System experiments are also expected to refine the PPN bounds. 

Among the most popular extensions to GR are scalar-tensor theories which introduce an additional scalar degree of freedom in the Einstein-Hilbert action that modifies the gravitational dynamics. 
Their linear cosmological behaviour is completely described by the EFT of dark energy, but their PPN description has not yet been fully developed.
This is predominantly due to the complexity arising from nonlinear screening mechanisms that are active on astrophysical scales.
These are an essential component of scalar-tensor theories, allowing significant modifications of GR at cosmological scales while still enabling them to satisfy stringent local constraints.
Developing the PPN formalism for Horndeski theories is therefore an important goal of current research.
For a comprehensive interpretation of observational constraints, one furthermore wishes to connect the cosmological and astrophysical formalisms and link the bounds inferred from the data in each regime.

In this paper, we developed the PPN formalism for Horndeski theories restricted to a luminal speed of gravitational waves and, as proposed in Ref.~\cite{Lombriser:2018guo}, we used the reconstruction of Horndeski models from the EFT of dark energy to establish a connection between these cosmological and astrophysical parameterisation frameworks.
The primary aim was to derive the corresponding PPN parameters.
Thereby we placed a particular emphasis on the general
incorporation of screening mechanisms.
Owing to their inherently nonlinear nature
they can be problematic when proceeding with the PN expansion as
its linearisation renders screening ineffective.
To tackle this, we employed the scaling method of Ref.~\cite{McManus:2016kxu} to first isolate the effective equations of motion in the screened region,
which then allowed the effects of the scalar field to be quantified in the PN approximation.
To this end, in Sec.~\ref{sec:einstein_gravity_limit} we applied the scaling method to
ensure that the Einstein field equations are recovered in the screened regions of the reconstructed Horndeski theories either through derivative or large field value screening. 
This led to a number of conditions on the powers of the scaling parameter that must be satisfied in order for a model to recover GR. 
Once the dependence of the perturbations on the scaling parameter was identified,
we could proceed with the expansion of the field equations in the parameter $\alpha$. 
As expected from screening, the models were found to recover GR at order $\alpha^0$.
We then identified the corrections to the Einstein field equations as well as the behaviour of the scalar perturbation at order $\alpha^q$. 
In Sec.~\ref{sec:PPN} we performed the PN expansion of the %
field equations at different orders in the scaling parameter to obtain 
the PN corrections.
Using these corrections we derived an expression for the effective gravitational constant as well as the PPN parameters $\gamma$ and $\beta$ by direct comparison with the PPN metric.
Interestingly, despite the fact that the scaling method depends on the screening mechanism, the form of the PPN parameters is universal.
Only the field equation for the scalar perturbation
and the screening potential $\Phi_{Sc}$ are theory dependent. 

A natural extension of this work is the application of our formalism to a range of popular scalar-tensor theories.
In particular, our framework allows one to translate the numerous experimental constraints on the PPN parameters
into bounds on the scaling parameter $\alpha$, given a normalisation of the scalar field background $\phi_0$, which can then be mapped into constraints on specific model parameters.
In Appendix~\ref{app:pulsar_test} we provide a preliminary example of how this can be accomplished. 
Under various assumptions, we use a recent upper bound on the Nordtvedt parameter, a linear combination of the PPN parameters~\cite{Nordtvedt:1968zz} inferred from a pulsar triple system to constrain a cubic Galileon and a $f(R)$ gravity model. 
However, we emphasise that a thorough analysis of the assumptions and the potential of such constraints will be the subject of future work. 
Another objective of future work
will be an analysis of how constraints placed on the PPN parameters translate into constraints on the EFT parameters at different orders in the EFT expansion or vice versa. 
As one can implement any type of screening mechanism while keeping the EFT functions unchanged, translating PPN constraints to the EFT parameters is nontrivial.
Finally, in Appendix~\ref{subsec:Einstein_limit_all} we demonstrate how the scaling approach can alternatively be performed at the level of the action instead of at the level of the field equations.
In future applications this may significantly simplify the procedure of determining whether a theory possesses an Einstein gravity limit, as well as identify the terms in the action that govern the screening effect.
In principle, this will allow one to directly read off the profile of the gravitational modifications in a given matter distribution from the action or conversely, determine an action for a desired profile.
A systematic combination of observational constraints from astrophysical and cosmological scales in a near model-independent manner is undoubtedly a challenging undertaking, but we are hopeful that the further development of our framework will significantly contribute to enabling progress towards this aim.

\section*{Acknowledgements}

This work was conducted in the context of C.R.’s MSc thesis at ETH Z\"urich.
C.R.~thanks Lavinia Heisenberg for helpful discussions and her and L.L.~for the great supervision.
C.R.~received the support of the Swiss Mobility programme for conducting work on his thesis at the University of Geneva. J.K.~and LL.~acknowledge the support of a Swiss National Science
Foundation Professorship grant (No.~170547).
Please contact the authors for access to research materials.


\appendix

\section{Useful relations}

For completeness and convenience to the reader we provide a summary of relations used in our derivations in Secs.~\ref{sec:background}--\ref{sec:PPN}.

\subsection{Field equations for Horndeski theories with $c_T=1$}\label{app:field_equations}

We write the trace-reversed metric and scalar field equations with the help of Refs.~\cite{McManus:2016kxu,Kobayashi:2011nu}. The metric field equations can be written as 

\begin{equation}
    G_4(\phi) R_{\mu\nu}=-\sum_{i=2}^{4}R^{(i)}_{\mu\nu}+\left(T_{\mu\nu}-\frac{1}{2}g_{\mu\nu}T \right)/M_p^{2} \,,
\end{equation}
and the scalar field equation is given by
\begin{equation}
G_4(\phi)\sums_{i=2}^4(\nabla^{\mu}J_{\mu}^{(i)}-P_{\phi}^{(i)}) 
+G_{4\phi} \sums_{i=2}^{4} R^{(i)}=-\frac{T}{M_p^{2}} G_{4\phi} \,,
\end{equation}
where 
\begin{align}
 P_{\phi}^{(2)}:=&G_{2\phi} \,, \\
 P_{\phi}^{(3)}:=&\nabla_{\mu}G_{3\phi}\nabla^{\mu}\phi\,, \\
  P_{\phi}^{(4)}:=&G_{4\phi}R\,,\\
  J^{(2)}_{\mu}:=&-G_{2X}\nabla_{\mu}\phi\,, \\
    J^{(3)}_{\mu}:=&G_{3X}\Box \phi\nabla_{\mu}\phi-G_{3X}\nabla_{\mu}X-2G_{3\phi}\nabla_{\mu}\phi\,, \\
     R^{(2)}_{\mu\nu}:=&-\frac{1}{2}G_{2X}\nabla_{\mu}\phi\nabla_{\nu}\phi-\frac{1}{2}g_{\mu\nu}\left(XG_{2X}-G_{2} \right)\,,\\
 R^{(3)}_{\mu\nu}:=&G_{3X}\left(\frac{1}{2}\Box\phi\nabla_\mu\phi\nabla_\nu\phi+\nabla_{(\mu}X\nabla_{\nu)}\phi+\frac{1}{2}g_\munu X\Box\phi\right)+G_{3\phi}\nabla_{\mu}\phi\nabla_{\nu}\phi\,, \\
 R^{(4)}_{\mu\nu}:=&-G_{4\phi}\left(\nabla_\mu\nabla_\nu\phi+\frac{1}{2}g_\munu \Box\phi\right)\,.
\end{align}

\subsection{Reconstructed Horndeski functions using $\{\alpha_M,\alpha_B,\alpha_K,\alpha_T\}$}\label{app:reconstruction}

For completeness we write the set of functions $\left\{\Omega,\Lambda,\Gamma,M_{2}^{4},\bar{M}_{1}^{3},\bar{M}_{2}^{2}\right\}$ with respect to the set $\{\alpha_M,\alpha_B,\alpha_K,\alpha_T\}$ using \cite{Kennedy:2017sof} (also see Refs.~\cite{Kennedy:2018gtx,Kennedy:2019nie,Kennedy:2020ehn}):
\begin{align}
    \Omega&=\frac{M^{2}}{M_{*}^{2}}c_{T}^{2}\,,\\
    \Gamma&=-\frac{\rho_{m}}{M_{*}^{2}}-\frac{M^{2}}{M_{*}^{2}}\beta \,,\\
    \Lambda&=\frac{M^{2}}{M_{*}^{2}}\left[3H^{2}c_{T}^{2}(1+\alpha_{M})+\beta+3H\dot{\alpha}_{T}\right],\\
    M_{2}^{4}&=\frac{1}{4}\rho_{m}+\frac{M^{2}}{4}\left[H^{2}\alpha_{K}+\beta \right]\,,\\
    \bar{M}_{1}^{3}&=M^{2}\left[H\alpha_{M}c_{T}^{2}+\dot{\alpha}_{T}-2H\alpha_{B}   \right]\,,\\
    \bar{M}_{2}^{2}&=-\frac{1}{2}M^{2}\alpha_{T}\,,
\end{align}
where $c_T^2=1+\alpha_T$ and 
\begin{align}
    \beta(t) \equiv c_{T}^{2}\left[2\dot{H}+H\dot{\alpha}_{M}+\alpha_{M}\left(\dot{H}-H^{2}+H^{2}\alpha_{M}\right)\right]+ H\dot{\alpha}_{T} (2\alpha_{M}-1)+\ddot{\alpha}_{T}\,.
\end{align}
Furthermore, we also write the functions $\xi^\I_n(\phi)$ as in Table~\ref{table:xi_n} with respect the the set of functions $\{\alpha_M,\alpha_B,\alpha_K\}$, where we set $\alpha_T=0$ and $G_4=tM_*^2$:
\begin{align}
    \xi_0^\2(t)=&-\frac{M_*^2}{2} \left[2 t \left(3 + 2 \alpha_M(t) + \alpha_M(t)^2\right) H(t)^2 \right.\nonumber\\
    &+ H(t) \left(2 M_*^2 \alpha_B(t) - M_*^2 \alpha_M(t) + t \left(4 + 2 M_*^2 \dot{\alpha}_B(t) - 
         M_*^2 \dot{\alpha}_M(t)\right)\right) \nonumber\\
   &\left.+t \left(2 M_*^2 \alpha_B(t) - \left(-2 + M_*^2\right) \alpha_M(t) + 
      2 \dot{\alpha}_M(t)\right) \dot{H}(t)\right]\,,
\\[1em]
    \xi_1^\2(t)=&\frac{1}{2} \left[M_*^2 t \left(-6 \alpha_B(t) + \alpha_M(t) \left(2 + \alpha_M(t)\right)\right) H(t)^2 + \rho_m(t) \right.\nonumber\\
    &+ H(t) \left(2 M_*^4 \alpha_B(t) - M_*^4 \alpha_M(t) + 
      M_*^2 t \left(2 + 2 M_*^2 \dot{\alpha}_B(t) - 
         M_*^2 \dot{\alpha}_M(t)\right)\right) \nonumber\\
         &\left.+ M_*^2 t \left(2 M_*^2 \alpha_B(t) + \alpha_M(t) - M_*^2 \alpha_M(t) + 
      \dot{\alpha}_M(t)\right) \dot{H}(t)\right]\,,
\end{align}
\begin{align}
    \xi_2^\2(t)=&\frac{1}{8} \left[M_*^2 t \left(6 \alpha_B(t) + \alpha_K(t) + \left(-4 + \alpha_M(t)\right) \alpha_M(t)\right) H(t)^2 + \rho_m(t) \right.\nonumber\\
    &+ H(t) \left(-2 M_*^4 \alpha_B(t) + M_*^4 \alpha_M(t) + 
      M_*^2 t \left(2 - 2 M_*^2 \dot{\alpha}_B(t) + 
         M_*^2 \dot{\alpha}_M(t)\right)\right) \nonumber\\
         &\left.+ M_*^2 t \left(-2 M_*^2 \alpha_B(t) + \left(1 + M_*^2\right) \alpha_M(t) + 
      \dot{\alpha}_M(t)\right) \dot{H}(t)\right]\,,
\\[1em]
    \xi_0^\3(t)=&-\frac{t}{2} \left(-2 \alpha_B(t) + \alpha_M(t)\right) H(t)\,,
\\[1em]
    \xi_1^\3(t)=&\frac{t}{2} \left(-2 \alpha_B(t) + \alpha_M(t)\right) H(t)\,,
\\[1em]
    \xi_0^\4(t)=&M_*^2\; t\,,
\\[1em]
    \xi_1^\4(t)=&0\,.
\end{align}

\subsection{Unscreened limit}\label{app:unscreened}

As explained in Ref.~\cite{McManus:2016kxu} it is also possible to recover the equations of motion of a scalar-tensor theory in the unscreened limit using the scaling method. 
We again distinguish between the different classes of screening mechanisms.
In large field value screening the screening mechanism is caused by a modification of the background value of the scalar field in a high-density region, translated into $\alpha\ll \psi$ and non-zero multiplicities. %
However, in derivative screening the mechanism 
is driven by a large coupling parameter, 
translated into $\alpha\gg\psi$ in our method. 
This difference impacts the method of finding the unscreened regime of a reconstructed theory.

We first consider derivative screening.
In the unscreened regime the coupling parameter should be small compared to the scalar field so we consider $\alpha\rightarrow 0$ as our unscreened limit. 
Other than taking the different limit, the method is equivalent to Sec.~\ref{subsubsec:derivative_screening_test}. 
As the set $Q_\m$ in Eq.~\eqref{eq:QM} is independent of the limit we consider we take it as the set of possible values for $q$. 
However, in this case we require its maximum value to avoid divergences in the metric field equation, namely $q=0$.
Furthermore, taking $Q_\s$ as in Eq.~\eqref{eq:QS}, the only values for $q$ allowed by the scalar field equation are also zero or negative. 
We are therefore left with the unique choice $q=0\iff 0\in Q_\s$. 
Note that with this choice we do not recover the Einstein equation in the unscreened region as expected.
To summarise, 
we expand the scalar field as $\phi=\phi_0(1+\psi)$, take the limit $\alpha\rightarrow 0$ and check that the scalar field equation is consistent in this limit.

When studying the chameleon mechanism in Sec.~\ref{subsubsec:exple_chameleon} or more explicitly in Ref.~\cite{McManus:2016kxu}, one realises that it is the value of the background field $\phi_0$ that has an impact on the screening mechanism. The background value of the field varies between the screened and the unscreened regions, $\phi_0\leftrightarrow\phi_0^*$ respectively.
In this method, it will be translated in a change of multiplicity for the functions $\zeta_{mk}^\I(\phi)$. 
Indeed as we saw in Sec.~\ref{subsubsec:large_field_value_test} when considering the screening limit $\alpha\rightarrow 0$, some of the multiplicities have to be non-zero at the background value $\phi_0$ in order to recover the Einstein field equations. 
However in the unscreened region, the background value of the field changes from $\phi_0$ to $\phi_0^*$ and the $\zeta$-functions have zero multiplicities for $\phi_0^*$. 
Now if one performs the scaling method for the limit $\alpha\rightarrow \infty$, one would find that we either recover the Einstein field equations or we have a divergence in the field equations, following the method for derivative screening. Therefore, this cannot be the unscreened limit for chameleon theories. 
If one considers the second limit $\alpha\rightarrow 0$ as our unscreened regime, one finds that the unique value for $q$ is zero, as in the derivative screening case. 
Thus, we can summarize the method as follows: to find the unscreened regime, we apply the scalar field expansion $\phi=\phi_0^*(1+\psi)$ and then we take the limit $\alpha\rightarrow 0$.

\subsection{Newtonian potentials for the PPN formalism}\label{app:NP}

Finally, we specify the Newtonian potentials of the PPN formalism given in Ref.~\cite{Will:2018bme} as well as the potentials depending on the scalar field as written in Ref.~\cite{McManus:2017itv}:
\begin{align}
\Phi_N&:=\int \frac{\rho'}{|x-x'|}\dd^3x'\,,&
\tilde{\Phi}_1^{(p,q+2)}&:= -\frac{1}{4\pi}\int \frac{\psi'^{(p,q)} v'^2}{|x-x'|^3}\dd^3x'\,,\nonumber\\
\Phi_1&:= \int \frac{\rho' v'^2}{|x-x'|}\dd^3x',&\tilde{\Phi}_2^{(p,q+2)} &:= \int \frac{\rho' \psi'^{(i)}}{|x-x'|}\dd^3x'\,,\label{eq:appendix_NP}\\
   \Phi_2&:= \int \frac{\rho' U'}{|x-x'|}\dd^3x'\,,&
   \mathcal{A}_\psi^{(p,q+2)} &:= -\frac{1}{4\pi}\int \frac{\psi'^{(i)}  [\overline{v}'\cdot(\overline{x}-\overline{x}')]^2 }{|\overline{x} - \overline{x}'|^5}\dd^3x'\,,\nonumber\\
   \Phi_3&:= \int \frac{\rho' \Pi'}{|x-x'|}\dd^3x',&\mathcal{B}_\psi^{(p,q+2)} &:= -\frac{1}{4\pi}\int \frac{\psi'^{(i)}  [\overline{a}'\cdot(\overline{x}-\overline{x}')] }{|\overline{x} - \overline{x}'|^3}\dd^3x'\,,\nonumber\\
   \Phi_4&:= \int \frac{p'}{|x-x'|}\dd^3x'\,.\nonumber
\end{align}
Some of these potentials can be found using the ordinary differential equations (ODEs) they satisfy. One can recover the ODEs with the rule 
\begin{align}
    \Phi=\int\frac{f(x')}{|x-x'|}\dd^3x'\implies \Delta\Phi=-4\pi f(x) \,.
\end{align}

\section{Scaling method at the level of the action}\label{subsec:Einstein_limit_all}

Up to this point we have been applying the scaling method at the level of the field equations.
One might expect however that it is possible to apply similar techniques 
directly at the level of the action. 
We briefly outline 
how this can be done by applying it to the same cubic Galileon and chameleon models discussed in Ref.~\cite{McManus:2016kxu}. 
Such a method could perhaps also allow an effective description of screening mechanisms constructed in a similar manner to EFT.
It would for instance allow to directly read off the radial dependence of the gravitational modification in the screened regime from a given action and matter distribution~\cite{Lombriser:2016zfz} that could be used in $N$-body simulations~\cite{Hassani:2020rxd}.

Starting from the action of a cubic Galileon,
\begin{align}
    S[\phi,g]=\frac{M_p^2}{2}\int\dd^4x\sqrt{-g}\bigg( \phi R+\frac{2\omega}{\phi}X-\alpha\frac{X}{4\phi^3}\Box\phi\bigg) \,,
\end{align}
we expand the scalar field into a background field and a perturbation as $\phi=\phi_0(1+\alpha^q\psi)$ 
resulting in
\begin{align}
    S=\frac{M_p^2\phi_0}{2}\int\dd^4x\sqrt{-g}\bigg( R+\alpha^q\psi R+2\omega\alpha^{2q}\tilde{X}-\alpha^{1+3q}\frac{\tilde{X}}{4\phi_0}\Box\psi\bigg) \,,\label{eq:new_scaling_action_cubic}
\end{align}
where $\tilde{X}$ is the kinetic term of the scalar perturbation $\psi$. 
In order to recover GR in the screened limit $\alpha\rightarrow \infty$ we require that all terms other than the Ricci scalar $R$ vanish.
Examining the action in Eq.~\eqref{eq:new_scaling_action_cubic} we see that this condition is satisfied for $q<-1/3$. 
The behaviour of the scalar field perturbation is described by the action at the order $\alpha^q$. 
Since the scalar field should be sourced by the energy-momentum tensor, the term $\alpha^q\psi R$ should be included in the action describing the scalar field perturbation. 
Furthermore, for consistency another term should be of order $\alpha^q$, otherwise the scalar field equation would be reduced to $R=T=0$.  
Thus we need to choose a value for $q$ such that $\alpha^q$ is second-to-leading order after $\alpha^0$ with at least two terms of order $\alpha^{q}$ including $\alpha^q\psi R$. 
This leads to the unique solution $q=-1/2$, in agreement with the analysis of Ref.~\cite{McManus:2016kxu} at the level of the field equations. 

We now consider the following model with chameleon screening,
\begin{align}
   S[\phi,g]=\frac{M_p^2}{2}\int d^4x \sqrt{-g}\left( \phi R +\frac{2\omega}{\phi}X - \alpha(\phi-\phi_0)^n \right)\,,
\end{align}
where we assume $n$ is positive for simplicity. 
After the scalar field expansion this becomes 
\begin{align}
 S=\frac{M_p^2}{2}\int d^4x \sqrt{-g}\left( \phi_0 R+\phi_0\alpha^q\psi R + 2 \omega\phi_0\alpha^{2q} \tilde{X} - \alpha^{1+nq}(\phi_0\psi)^n \right) \,.
\end{align}
We follow a similar argument to derivative screening but now taking the limit $\alpha\rightarrow 0$. 
In order to recover the Einstein-Hilbert action in this limit we require $q>0$.
In addition, to have two terms of order $\alpha^q$ we find that $q=1/(1-n)$. 
From these two conditions we conclude that a consistent Einstein gravity limit exists only when $n<1$, 
in agreement with the previous analysis at the level of the field equations. 

In order to extend this method to 
more general Horndeski theories with
$G_4(\phi,X)\neq \phi$ we require
the perturbed action to include a term proportional to the Ricci scalar so that $aR$ with $a$ constant and a term of the form $f(\psi)R$ which sources the scalar field with the stress-energy tensor. 
We leave the extension to general reconstructed Horndeski theories to future work.

\section{Application: Towards pulsar tests of the cubic Galileon and $f(R)$ gravity}\label{app:pulsar_test}

Finally, we shall provide here an exploratory analysis of PPN constraints that can be inferred employing our formalism.
We stress however that this analysis will only serve as an initial estimate for such constraints.
A more detailed analysis, accounting more carefully for the complex observational and phenomenological aspects involved will be subject to future work.
We will focus on pulsar systems, which have proven to be great testing grounds for gravitational theories.
The radio pulses that neutron stars emit are timed very precisely and they allow us to deduce the orbital behaviour of these systems with significant precision. 
In particular, the discovery of the triple pulsar system PSR J0337+1715 \cite{Ransom:2014xla}, a neutron star accompanied with two white dwarfs, sets the strongest constraint on the Nordtvedt parameter $\abs{\eta}<2.6\cdot 10
^{-5}$~\cite{Archibald:2018oxs}. 
It was shown in Ref.~\cite{Nordtvedt:1968zz} that the trajectory of freely falling objects will depend on their compactness, when $\eta\neq 0$. 
For Horndeski theories, this parameter is of the form
\begin{align}
\eta=4\beta-\gamma-3 \,.
\end{align}
The triple system is composed of a pulsar with an inner companion orbiting close by and a second one orbiting further away. 
The upper bound on the parameter was set by observing the relative motion of the inner system in the gravitational field of the outer companion. 
We will now briefly discuss how the PPN formalism developed in this work can be used with the current constraint on the Nordtvedt parameter. 
We apply the formalism to the cubic Galileon~\cite{Chow:2009fm} as well as Hu-Sawicki $f(R)$ gravity~\cite{Hu:2007nk} and we use the orbital data of the triple system that can be found in Ref.~\cite{Archibald:2018oxs}.

In order to simplify the calculations we make the following assumptions: 
(i) we assume circular orbits, which means that the inner system is at constant distance from the outer white dwarf; 
(ii) we consider the inner orbit to be small compared to the outer orbit; 
(iii) we take $\Pi=0$; 
(v) the outer white dwarf is modeled by a perfect fluid with constant density $\rho$; 
(vi) pressure in the outer white dwarf is found using the Tolman-Oppenheimer-Volkoff equation~\cite{Straumann:2013spu}; 
(vii) all velocities in the system are negligible; 
(viii) we assume smoothness of the PPN potentials between inside and outside the outer white dwarf. 
In addition, Newton's constant $G_N$ is constrained from Solar System experiments where the scalar field is screened. 
Hence we can relate the effective gravitational constant $G_\eff$ in Eq.~\eqref{eq:Geff_PPN} and the Newton's constant using $G_\eff\cong G/\phi_0=G_N$. 
We normalise the background value of the scalar field to $\phi_0=1$.

Starting with the cubic Galileon theory with an action of the form 
\begin{align}
    S[g,\phi]=\frac{M_p^2}{2}\int\dd^4x\sqrt{-g}\left(\phi R-\alpha\frac{X^2}{\phi^4}-\frac{\alpha}{2\phi^3}X\Box\phi\right)+S_m[g] \, ,
\label{eq:action_galileon_cosmology}
\end{align}
where the scaling parameter $\alpha$ can be related to the usual cubic Galileon coupling parameter $\Lambda$ \cite{Joyce:2014kja} via $\alpha=\Lambda^{-3}$.
Using the discussion of Sec.~\ref{subsubsec:derivative_screening_test} or Appendix~\ref{subsec:Einstein_limit_all}, we find the order of the correction terms to be $\alpha^{q}$ with $q=-1/2$. 
To find the PPN potentials \eqref{eq:appendix_NP}, we need the equations for the scalar perturbation $\psi$ and the potential $\Phi_{Sc}$. 
Using Eqs.~\eqref{eq:scalar_eqt_lowest_order} and \eqref{eq:PDE_Phi_Sc} together with the value for $q$, we find
\begin{align}
    -16\pi G\rho&=\nabla_\mu\nabla_\nu\psi\nabla^\mu\nabla^\nu\psi-(\Box\psi)^2 \,,\\
    \nabla^{2} \Phi_\textrm{Sc}&=-\frac{1}{2}(\grad\psi)^2 \nabla^{2} \psi \,,
\end{align}
where $\rho$ is constant in the outer white dwarf and zero otherwise. 
The covariant derivatives of the scalar field equation can be replaced by $\nabla_\mu\rightarrow \partial_\mu$ and partial time derivatives $\partial_0$ are neglected in the PN limit. 
Under these assumptions as well as the measured distance between the pulsar and the outer companion, the PPN potentials and the scalar perturbation $\psi$ can be calculated for the inner system. 
In particular, the spherical symmetry considerably simplifies the PDEs and most of the potentials vanish, namely $\Phi_1=\Phi_3=\tilde{\Phi}_1=\mathcal{A}_\psi=\mathcal{B}_\psi=0$.
One can now relate the Nordtvedt parameter $\eta(\alpha)$ to the scaling parameter $\alpha$ through the PPN parameters $\gamma$ and $\beta$ in Eqs.~\eqref{eq:gamma} and \eqref{eq:beta}, respectively, so that a constraint on $\alpha$ can be found. 
By connecting the cubic Galileon parameter to the Vainshtein radius $r_V$ 
which describes the distance from the source beyond which there is no screening, 
we find the lower bound $r_V>3\cdot 10
^{-3}\,$pc. 
In comparison, the expected value of $r_V$ to satisfy cosmological observations is of order $r_V\sim 10^{4}\,$pc \cite{deRham:2014zqa}. 
Comparing our result to previous constraints obtained from pulsar observations, we find that it is of the same order of magnitude than what can be inferred from the multipole radiation of binary systems \cite{deRham:2012fw}.

Next we consider Hu-Sawicki $f(R)$ models~\cite{Hu:2007nk}, which are endowed with the chameleon mechanism. 
It was shown in Ref.~\cite{Chiba:2003ir} that $f(R)$ gravity is equivalent to a class of Brans-Dicke theories and so we can write the action of the
model as
\begin{align}
     S[\phi,g]=\frac{M_p^2}{2}\int\dd^4x\sqrt{-g}\left(\phi R-\alpha(\phi-1)^{n/(n+1)}\right)
\end{align}
with $\alpha$ as the scaling parameter and we choose $n=1$ for this example. 
Following the discussion in Sec.~\ref{subsubsec:large_field_value_test} or Appendix~\ref{subsec:Einstein_limit_all} we find the order of the scalar field and metric perturbations to be $\alpha^q$, with $q=2$. 
Eqs.~\eqref{eq:scalar_eqt_lowest_order} and \eqref{eq:PDE_Phi_Sc} give us the following relations for the theory dependent variables $\psi$ and $\Phi_{Sc}$,
\begin{align}
    \psi&=\left(\frac{M_p^2}{2\rho}\right)^{2} \,,\\
    \nabla^{2}\Phi_\textrm{Sc}&=-\sqrt{\psi} \,.
\end{align}
Contrary to the cubic Galileon case, we cannot take $\rho=0$ outside the outer white dwarf since it would lead to a division by zero for the scalar field.
This is a consequence of the environmental dependence in chameleon models.
We hence describe the ambient density with the Milky Way dark matter halo as in Ref.~\cite{Lombriser:2013eza}. 
From the position of the triple system in our sky and its distance from the Solar System~\cite{Ransom:2014xla} we determine its distance from the galactic centre to find the matter density. 
Using Appendix~\ref{app:NP} we find the value of the PPN potentials for the inner system and using the PPN parameters $\gamma $ and $\beta$ in Eqs.~\eqref{eq:gamma} and \eqref{eq:beta} respectively, we find the relation $\eta(\alpha)$. 
The physical parameter $f_{R0}\equiv df/dR|_{R=\bar{R}_0}$ that is usually constrained in Hu-Sawicki models shall represent here the background value of the scalar perturbation in the intergalactic medium around the Milky Way, taken to be the cosmological background, which is not quite accurate but a usual convention.
It is connected to the scaling parameter through $\alpha=2f_{R0}^{1/2}\bar{R}_0$, where $\bar{R}_0=3H_0^2(1+3(1-\Omega_m))$~\cite{Lombriser:2013eza} is the background curvature of space. 
Using the data from Ref.~\cite{Aghanim:2018eyx}, the bound on $\eta$ and its relation to $\alpha$, we find the constraint $\abs*{f_{R0}}<7 \times 10^{-4}$. 
Importantly, this upper bound is set by the galaxy density surrounding the pulsar system. 
In the chameleon mechanism, gravitational objects may become self-screened if their density is large enough, which means that they can become unaffected by the ambient scalar field, making this bound invalid.
However, in the case of the pulsar system, even though white dwarfs are compact objects,
the system we consider is only screened
for $f_{R0}\leq 2 \times 10^{-4}$ (following analogous arguments used for the Solar System in Ref.~\cite{Hu:2007nk}) such that this bound applies.
In contrast, Solar System experiments yield a constraint of $\abs*{f_{R0}} \lesssim 10^{-7}$~\cite{Hu:2007nk,Lombriser:2013eza,Lombriser:2014dua}, which is however set by the requirement that its location in the Milky Way should be screened by the galactic density
with the Sun itself being screened already for $f_{R0}\sim 10^{-2}$~\cite{Hu:2007nk}.
In that sense the pulsar constraint can be interpreted as being stronger than that set by the Sun.

\bibliographystyle{JHEP}
\bibliography{main.bbl}

\end{document}